\documentclass[a4paper,11pt]{article}
\usepackage{amsmath, amssymb, amsfonts,indentfirst,graphicx,color,anysize}
\marginsize{2cm}{2cm}{2cm}{2cm}
\title{
	\includegraphics[width=0.35\textwidth]{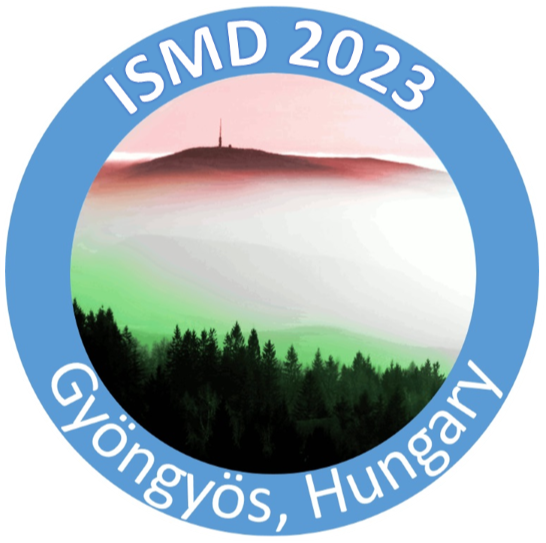}\\[1cm]
	\textbf{
QED meson description of the anomalous particles at $\sim$17 and $\sim$38 MeV\footnote{ Based on a talk presented at 52nd International Symposium on Multiparticle Dynamics,
at Gy\" ongy\" os, Hungary, August  20-26, 2023.
 URL: {{ https://indico.cern.ch/event/1258038/contributions/5538335/}}.}}
} 
\author{{Cheuk-Yin Wong}\\[1ex]
 Physics Division, Oak Ridge National Laboratory\footnote{
 {{The research was supported  in part by
  UT-Battelle, LLC, under contract DE-AC05-00OR22725 with the US
  Department of Energy (DOE). The US government retains the
  publisher, by accepting the article for publication, acknowledges
  that the US government retains a nonexclusive, paid-up, irrevocable,
  worldwide license to publish or reproduce the published form of this
  manuscript, or allow others to do so, for the US government
  purposes. DOE will provide public access to the results of
  federally sponsored research in accordance with the DOE Public
  Access Plan (http://energy.gov/downloads/doe-public-access-plan),
  Oak Ridge, Tennessee 37831, USA}}}, 
 Oak Ridge, Tennessee USA\\
}

\begin{document}
\def\bb    #1{\hbox{\boldmath${#1}$}}
 \def\oo    #1{{#1}_0 \!\!\!\!\!{}^{{}^{\circ}}~}  
 \def\op    #1{{#1}_0 \!\!\!\!\!{}^{{}^{{}^{\circ}}}~}
\def\blambda{{\hbox{\boldmath $\lambda$}}} 
\def\eeta{{\hbox{\boldmath $\eta$}}}
\def\bxi{{\hbox{\boldmath $\xi$}}} 
\def\bzeta{{\hbox{\boldmath $\zeta$}}}
\def\sD{D \!\!\!\!/}
\def\sd{\partial \!\!\!\!/}
\def\qcd{{{}^{\rm QCD}}}   
\def\qed{{{}^{\rm QED}}}   
\def\2d{{{}_{\rm 2D}}}         
\def\4d{{{}_{\rm 4D}}}         

\def\qcdu{{{}_{ \rm QCD}}}   
\def\qedu{{{}_{\rm QED}}}   
\def\qcdd{{{}^{ \rm QCD}}}   
\def\qedd{{{}^{\rm QED}}}   
\maketitle 
\vspace{-0.5cm}
\begin{abstract} 
The Schwinger confinement mechanism stipulates that a massless fermion
and a massless antifermion are confined as a massive boson when they
interact in the Abelian QED interaction in (1+1)D. If we approximate
light quarks as massless and apply the Schwinger confinement mechanism
to quarks, we can infer that a light quark and a light antiquark
interacting in the Abelian QED interaction are confined as a QED meson
in (1+1)D.  Similarly, a light quark and a light antiquark interacting
in the QCD interaction in the quasi-Abelian approximation will be
confined as a QCD meson in (1+1)D.  The QED and QCD mesons in (1+1)D
can represent physical mesons in (3+1)D when the flux tube radius is
properly taken into account.  Such a theory leads to a reasonable
description of the masses of $\pi^0, \eta$, and $ \eta'$, and its
extrapolation to the unknown QED sector yields an isoscalar QED meson
at about 17 MeV and an isovector QED meson at about 38 MeV.  The
observations of the anomalous soft photons, the hypothetical X17
particle, and the hypothetical E38 particle bear promising evidence
for the possible existence of the QED mesons.  Pending further
confirmation, they hold important implications on the properties on
the quarks and their interactions.
\end{abstract}

\newpage

\section{Introduction}

In the International Symposium on Multiparticle Dynamics in 2009 at
Gomel, Belarus, Perepelitsa reviewed the phenomenon of the anomalous
soft photons~\cite{Per09} and reported that in high-energy
hadron--hadron~\cite{Chl84,Bot91,Ban93,Bel97,Bel02pi,Bel02} and
$e^+e^-$ collisions~\cite{Per09,DEL06,DEL08,DEL10}, anomalous soft
photons in the form of excess $e^+e^-$ pairs, are produced at a rate
exceeding the standard model bremsstrahlung predictions by an average
factor of about four.  In particular, in exclusive DELPHI
hadron production measurements in $e^+e^-$ collisions at the $Z^0$
energy, the anomalous soft photons are proportionally produced
whenever hadrons are produced, but they are not produced when hadrons
are not produced~\cite{DEL06,DEL08}.  The transverse momenta of the
anomalous soft photons lie in the region of many tens of MeV/c,
corresponding to the production of neutral bosons with masses in the
region of many tens of MeV.  Many different descriptions have been put
forth to interpret the anomalous soft photons, including a cold quark
gluon plasma~\cite{Van89,Lic90,Lic94,Kok07}, pion condensate~\cite{Bar89},
pion reflection~\cite{Shu89}, corrections to bremsstrahlung~\cite{Bal90}, color flux tube particle production~\cite{Czy94},
stochastic QCD vacuum~\cite{Nac94,Leb22}, ADS/CFT supersymmetric
Yang--Mills theory~\cite{Hat10}, Unruh radiation~\cite{Dar91}, closed
quark--antiquark loop~\cite{Sim08}, QED-confined $q\bar q$ states~\cite{Won10,Won11,Won14,Won20,Won22,Won22a,Won22b,Won22c,Kos22,Won23},
and induced currents in the Dirac sea~\cite{Kha14}.

It was soon realized~\cite{Won10} that the simultaneous and
proportional production of the anomalous soft photons with hadrons
suggests that a parent particle of an anomalous soft photon is
likely to contain some elements of the hadron sector, such as a light
quark and a light antiquark.  Quarks and antiquarks carry color and
electric charges, and they interact with the QCD and the QED
interactions.  The parent particle of an anomalous soft photon
cannot arise from the light quark and the light antiquark interacting
non-perturbatively in the QCD interaction because such a
non-perturbative QCD interaction will endow the pair with a mass much
greater than the mass scale of the anomalous soft photons, in a
contradictory manner.  We are left only with the possibility of the
quark and the antiquark interacting non-perturbatively in the QED
interaction.  Such a possibility is further reinforced by the special
nature of attractive confining gauge interactions in which the smaller
the coupling constant, the lower the mass of the confined
composite particle. This is in contrast to non-confining attractive Coulomb-type or
Yukawa-type interactions for which the smaller the coupling constant,
the greater the mass of the composite system.  Furthermore,
the Schwinger confinement mechanism stipulates that a massless charged
fermion interacting with an antifermion in a gauge field with the
coupling constant $g_\2d$ in (1+1)D leads to a confined boson with a
mass~\cite{Sch62,Sch63}
\begin{eqnarray}
m_{\rm boson}=\frac{g_\2d}{\sqrt{\pi}},
\label{eq1}
\end{eqnarray}
 indicating that the mass of the confined fermion--antifermion
 composite system  is directly proportional to the coupling constant.
 Application of the Schwinger confinement mechanism to quarks
 interacting in the QED interaction will bring the quantized masses of
 a $q\bar q$ pair to the lower mass region of the anomalous soft
 photons.  It was therefore proposed in 2010~\cite{Won10} that a light
 quark and a light antiquark interacting non-perturbatively with the
 QED interaction may lead to new open-string QED boson states
 (QED-meson states) with a mass of many tens of MeV. These QED mesons
 may be produced simultaneously with the QCD mesons in the string
 fragmentation process in high-energy collisions, and the excess
 $e^+e^-$ pairs may arise from the decays of these QED mesons.  For
 quarks with two flavors in the massless quark limit, the masses of
 the isoscalar and isovector QED mesons were predicted to be 12.8 MeV
 and 38.4 MeV, respectively (Table I of~\cite{Won10}).

In a series of experiments in search of axions, Krasznarhorkay and
collaborators studied the $e^+ e^-$ spectrum in low-energy
proton fusion of light $\alpha^n$ nuclei with $n$ = 1, 2, \linebreak  and 3.  Since
2016, they have been observing the occurrence of the hypothetical neutral
``X17'' boson with a mass of about 17 MeV from the $e^+ e^-$
spectrum (i) in the decay of the 18.15 MeV $I(J^\pi)$=$0(1^+) $
excited $^8$Be state to the $^8$Be ground state~\cite{Kra16}, (ii) in
the decay of the 18.05 MeV $I(J^\pi)$=$0(0^-)$ excited $^4$He state to
the $^4$He ground state~\cite{Kra19,Kra21}, (iii) in the decay of the
off-resonance excited $^8$Be states to the $^8$Be ground state
\cite{Sas22}, and (iv) in the decay of the 17.23 MeV
$I(J^\pi)$=$1(1^-)$ excited $^{12}$C state to the $^{12}$C ground
state~\cite{Kra22}. Updates of the ATOMKI measurements on the
hypothetical X17 particle have also been presented
\cite{X1722,X17Kra,Kra23}.  In a recent measurement in the decay of
the $^8$Be 18.15 MeV state to the $^8$Be ground state in the
proton fusion on $^7$Li, the Hanoi University of Science reported the
observation of a significant structure, which indicates a hypothetical
neutral boson with a mass of about 16.7 MeV decaying into $e^+e^-$, in
support of the earlier ATOMKI observation~\cite{Tra23}.  An indirect
support for the hypothetical X17 particle comes from the $p_T$
spectrum of the anomalous soft photons in $pp$ collisions at $p_{\rm
  lab}$= 450 GeV/c~\cite{Bel02}, in the thermal model of the
transverse momentum distribution~\cite{Won20,Dya21}.

The ATOMKI observation of the hypothetical X17 particle has generated
a great deal of interest~\cite{Won10,Won11,Won14,Won20,Won22,Won22a,Won22b,Won22c,Kos22,Kra19,Nag19,Kra21,Sas22,Kra22,
X1722,X17Kra,Kra23,Tra23,Dya21,Zha17,Alv18,Fen16,Fen17,Fen20,Bat15,For17,Bor19,Cha22,Bor22,Bor23,Ros17,Ros19,Ros19a,Kub22,Ell16,Ban18,Ban18a,Ban20,Pad18,Viv22,Viv22a,Bar23,Var24}.
Although the mass of the hypothetical X17 particle was close to the
isoscalar QED meson predicted earlier in~\cite{Won10}, the X17 boson
led to many speculations as discussed in the Proceedings of the
Workshop on ``Shedding lights on the X17''~\cite{X1722}.  The proposed
models include the QED meson~\cite{Won10,Won11,Won14,Won20,Won22,Won22a,Won22b,Won22c,Kos22}, the
axion~\cite{Alv18}, the fifth force of Nature~\cite{Fen16}, a dark
photon~\cite{Bat15}, new physics particles~\cite{For17}, the Framed
Standard \mbox{Model~\cite{Bor19,Cha22,Bor22,Bor23},} Higgs doublet~\cite{Ros17},
a 12-quark state~\cite{Kub22}, a light pseudoscalar~\cite{Ell16}, and dressed QED radiation~\cite{Var24}.
The experimental confirmation of the hypothetical X17 particle is
being actively pursued by many laboratories~\cite{X1722}, including
ATOMKI~\cite{X17Kra,Kra23}, HUS~\cite{Tra23}, Dubna~\cite{Abr23}, New JEDI~\cite{JEDI}, STAR~\cite{x17STAR}, MEGII~\cite{x17MEG, MEGIIISMD23}, ATLAS~\cite{ATLASISMD23,ATLAS2ISMD23}, CTU Prague~\cite{x17Prague,Luz23},
NTOF~\cite{x17NTOF}, NA64~\cite{x17NA64}, INFN-Rome~\cite{x17INFNRome}, NA48~\cite{x17NA48}, Mu3e~\cite{x17Mu3e},
MAGIX/DarkMESA~\cite{x17MAGIX}, JLAB PAC50~\cite{x17JLAB,x17JLAB1},
PADME~\cite{x17PADME,Rag23}, DarkLight~\cite{ARIEL,Tre22}, LUXE~\cite{Hua22}, FASER~\cite{Fen23}, ANU/UM~\cite{Kib23},\linebreak and   Montreal~\cite{Mon23}.

In separate experiments, Abraamyan and collaborators at Dubna have
been using the two-photon decay of a neutral boson to study the
resonance structure of the lightest hadrons near their energy thresholds~\cite{Abr09}.  Upon the suggestion of van Beveran and \mbox{Rupp~\cite{Bev11,Bev12,Bev12a,Bev20}, }the Dubna Collaboration undertook a search for the E38
particle in $d$(2.0~GeV/n) + C, $d$(3.0~GeV/n) + Cu and $p$(4.6~GeV) + C
reactions with internal targets at the JINR Nuclotron.  They observed
that the invariant masses of the two-photon distributions exhibit a
resonance structure at around 38 MeV~\cite{Abr12,Abr19}.  In a recent
analysis in the diphoton spectrum in the lower invariant mass region,
the Dubna Collaboration reported the observation of resonance-like
structures both at $\sim$17 and $\sim$38 MeV~\cite{Abr23}, in support
of the earlier ATOMKI observation of the hypothetical X17 particle and the
earlier Dubna observation of the hypothetical E38 particle~\cite{Abr12,Abr19}. An indirect supporting signal for the
hypothetical E38 particle comes from the $p_T$ spectrum of anomalous
soft photons in $e^+e^-$ annihilations at the $Z^0$ resonance energy
of $\sqrt{s} = 91.18 $ GeV, which is consistent with the production of a
neutral boson with a mass of about 38 MeV, in the thermal model of the
transverse momentum distribution~\cite{Won20,Dya21}.
 
While there are many different theoretical interpretations, the open
string $q\bar q$ QED meson model mentioned above~\cite{Won10,Won20}
holds the prospect of describing the anomalous particles in a
consistent framework.  We would like to review here how such a model
of QED and QCD mesons emerges as a reasonable theoretical concept
consistent with experimental observations.  We would also like to
examine the implications for the existence of the QED mesons, if they
are confirmed by future experimental measurements.

It is worth pointing out that the cross-fertilization between condense
matter physics and particle physics brings bountiful fruits on the
physics frontiers.  In this respect, the confinement of electric
charges and electric anticharges (Cooper pairs and anti-Cooper pairs
in a specific case) in compact QED due to a linear potential
(confining string) has been experimentally observed in condensed
matter systems, where it gives rise to a new state of matter called
super-insulators\footnote{The Author thanks the Referee of the Special
Issue of Universe for bringing this to his attention.}
(see~\cite{Dia18,Dia18a,Dia20,Dia20a,Dia21}).  There are similarities
and also differences of possible quark confinement of QED mesons in
QED in particle physics and charge-anticharge confinement in QED in
condense matter physics as one can surmise by comparing Figure~1 of
ref.~\cite{Dia18} and Figure~2b of ref.~\cite{Won23}.  Whereas the
confinement of QED mesons examined here is concerned with the QED
confinement of massless fermions, the confinement of (charged
boson)--(anticharged boson) in super-insulators in condenser matter
physics is concerned with the QED confinement of massive bosons.  The
experimental existence of the QED super-insulators supports the
possible existence of the QED mesons, which are good candidates for
the X17 and E38 particles.  Future parallel investigations on the
common question of charge confinement in the QED interaction will
bring benefits to both fields.

 \section{The Schwinger Confinement Mechanism}

Schwinger showed in 1963 that a massless fermion and its antifermion
interacting in the Abelian U(1) QED gauge interaction are bound and
confined as a neutral QED boson with a mass~\cite{Sch62,Sch63} as
given by Equation (1) where the coupling constant $g_\2d$ in (1+1)D has
the dimension of a mass.  Such a Schwinger confinement mechanism
occurs for massless fermions interacting in Abelian U(1) gauge
interactions of all strengths, including the interaction with a weak
coupling (as in QED), as well as the interaction with a strong
coupling (as in QCD in the quasi-Abelian QCD approximation, which we
shall introduce~below).
 
The masses of light quarks are about a few MeV~\cite{PDG19,PDG22}.
Therefore, light quarks can be approximated as massless fermions, and
we can apply the above Schwinger mechanism for massless fermions to
quarks and antiquarks interacting in the QED interaction in (1+1)D.
By such an application, we infer that a light quark and a light antiquark
are bound and confined as a neutral QED boson in (1+1)D.  From the
works of Coleman, Jackiw, and Susskind~\cite{Col75,Col76}, we can
infer further that the Schwinger confinement mechanism persists even
for massive quarks in (1+1)D.

It is instructive to review the Schwinger confinement mechanism here
to understand how a light quark $q$ and a light antiquark $\bar q$
approximated as massless can be confined in QED in (1+1)D.  From the
electrostatic viewpoint, the electric lines of force in (1+1)D
originate from the positive quark at $x_q$ to end at the negative
antiquark at $x_{\bar q}$, and the quarks experience a confining
interaction~\cite{Col76}:
\vspace{-6pt}\begin{eqnarray}
V_{q\bar q}=\frac{g_\2d^2}{2} |x_q - x_{\bar q}|.
\end{eqnarray}

Such a confining interaction is one of the causes of quark confinement
in (1+1)D.   The spectrum of such a confining potential has been calculated 
in Appendix D of~\cite{Won22}. 
 For light quarks which can be approximated as massless,
there is an additional dynamical quark effect beyond the static linear
interaction alone.  The dynamics of quark matter current $j^\mu$
depends on the interacting gauge field $A^\mu$, which in turn depends
on the quark current $j^\mu$ in return, in an infinite loop of
matter($j^\mu$)-and-field($A^\mu$) interaction that facilitates the
confinement both of the matter field ($j^\mu$) and the gauge field
($A^\mu$).  As a consequence, neither the quark field nor the gauge
field make their external appearances, and there exists only a neutral massive boson field
 which emerges as free boson quanta containing both the
quark field and the gauge field.

We consider the interacting quark--QED system as a quark--QED fluid and
envisage the vacuum of the interacting quark--QED fluid as a calm Dirac
sea with the lowest-energy state to consist of quarks filling up the
(hidden) negative-energy Dirac sea and to interact with the QED
interaction in (1+1)D space-time with coordinates $x=(x^0,x^1)$.  The
quark--QED vacuum is defined as the state that contains no valence
quarks as particles above the Dirac sea and no valence antiquarks as
holes below the Dirac sea.  Subject to the applied disturbing gauge
field $A^\mu(x)$ with a coupling constant $g_\2d$ in (1+1)D, the
massless quark field $\psi(x)$ satisfies the Dirac equation,
\begin{eqnarray}
\gamma_\mu [ p^\mu - g_{\2d} A^\mu(x)] \psi(x) =0.
\label{eq7}
\end{eqnarray}
 
 The applied gauge field $A^\mu(x)$ governs the motion of the quark
 field $\psi(x)$.  From the motion of the quark field $\psi(x)$, we
 obtain the induced quark current $j^\mu(x)=\langle \bar \psi (x)
 \gamma^\mu \psi(x) \rangle$.  If we consider only the sets of states
 and quark currents that obey the gauge invariance by imposing the
 Schwinger modification factor to ensure the gauge invariance of the
 quark Green's function, the quark current $j^\mu(x) $ at the
 space-time point $x$ induced by the applied $A^\mu(x)$ can be
 evaluated.  After the singularities from the left and from the right
 cancel each other, the gauge-invariant induced quark current $
 j^\mu(x)$ is found to relate explicitly to the applied QED gauge
 field $A^\mu(x)$ by~\cite{Sch62,Sch63,Won94}
\vspace{-6pt}\begin{eqnarray}
j^\mu (x)  = -\frac{g_\2d}{\pi }\left ( A^\mu  (x) - \partial ^\mu \frac{1}{\partial _\eta \partial ^\eta} \partial _\nu A ^\nu (x) \right ).
\label{eq4}
\end{eqnarray}

We can understand the first term on the right-hand side of the above
equation intuitively as indicating that the induced current
gains in strength as the strength of the applied gauge field
$A^\mu(x)$ increases, and it acquires a sign opposite to the sign of
the applied gauge field $A^\mu(x)$ because an electric charge attracts
charges of the opposite sign.  The additional term on the right-hand
side consists of the linear function of $A^\mu(x)$ and a particular 
functional combination of partial derivatives that ensures the
gauge invariance between $j^\mu(x)$ and $A^\mu(x)$.  Upon a change of
the gauge in $(A^\mu)' (x) \to A^\mu(x) - \partial ^\mu \Lambda(x)$
for any local function of $\Lambda(x) $, the gauge invariance of the
relationship between $j^\mu(x)$ and $A^\mu(x)$ can be easily
demonstrated by direct substitution in Equation~(\ref{eq4}).

The quark current $j^\mu(x)$ in turn generates a new gauge field
$\tilde A^\mu(x)$ through the Maxwell equation,
\begin{eqnarray}
\hspace*{-0.3cm}\partial_\nu F^{\nu \mu}(x) =\partial_\nu  \{ \partial^\nu \tilde A^\mu (x)  -  \partial^\mu \tilde A^\nu (x) \}=g_\2d j^\mu (x) = g_\2d \langle \bar \psi (x) \gamma^\mu \psi (x) \rangle .
\label{eq5}
\end{eqnarray}

A stable collective particle--hole excitation of the quark system
occurs when the initial applied $A^\mu(x)$ gives rise to the induced
quark current $j^\mu$, which in turn leads to the new gauge field
$\tilde A^\mu(x)$ self-consistently.  We impose this self-consistency
condition of the gauge field, $A^\mu(x)=\tilde A^\mu(x) $.  In that
case, Equations~(\ref{eq4})~and~(\ref{eq5}) are mathematically the same as
\vspace{-6pt}\begin{eqnarray}
\hspace*{-0.8cm}
&&\partial _\nu \partial ^\nu A^\mu (x) + \frac{g_\2d^2}{\pi }A^\mu(x) =0, 
\\
\hspace{-6.0cm}{\rm and~}~~~~~~~~~~~~~~~~~~~~~~~~~~~~
&&\partial _\nu \partial ^\nu j^\mu (x) +\frac{g_\2d^2}{\pi }j^\mu (x) =0.\!
\end{eqnarray}

{We} 
can follow Casher, Kogut, and Susskind~\cite{Cas74} and also
Coleman et~al.~\cite{Col75,Col76} to introduce a boson field
$\phi(x)$ related to the current $j^\mu(x) $ by
\begin{eqnarray}
j^\mu (x) = \epsilon^{\mu \nu} \partial _\nu  \phi (x),
\end{eqnarray}
where $\epsilon^{\mu \nu}$ is the antisymmetric unit tensor with
$\epsilon^{01} = \epsilon^{10}$ = 1.   The boson field $\phi$, the quark current 
$j^\mu$,  and the gauge field $A^\mu$ are all related to each other, and thus, the boson field $\phi(x)$  also satisfies 
the Klein--Gordon equation,
\vspace{-6pt}\begin{eqnarray}
&&\partial _\nu \partial ^\nu \phi(x)   + \frac{g_\2d^2}{\pi }\phi (x) = 0, 
\end{eqnarray}
with a boson mass  given by {Equation} (1).~
Hence, a massless quark and a massless  antiquark interacting in the QED gauge
interaction in (1+1)D  form a bound and confined neutral QED meson $\phi$ 
with the mass  $m=g_\2d/\sqrt{\pi}$.  
 
The QED-confined meson in (1+1)D can be viewed in two equivalent ways
\cite{Won10,Won20}.  It can be depicted effectively as a QED-confined
one-dimensional open string, with a quark and an antiquark confined at
the two ends of the open string subject to an effective linear
two-body confining interaction.  A more basic and physically correct
picture depicts the QED meson as the macroscopic manifestation of the collective space-time oscillation  of the gauge field
$A^\mu(x)$ associated with  the microscopic
particle--hole excitation of quarks from the Dirac sea.
Through the coupling 
of    the gauge field $A^\mu(x)$ to the quarks  
in the Dirac equation, a space-time variation
of the gauge field $A^\mu(x)$ leads to a space-time variation in the
 quark current $j^\mu(x)$, which in turn determines the space-time
variation of the gauge field $A^\mu(x)$ through the Maxwell equation~\cite{Sch62,Sch63,Won94}.  As a consequence of such a
self-consistent coupling, a quantized and locally confined space-time
collective variations of the macroscopic  QED gauge
field $A^\mu(x)$ can sustain themselves indefinitely at the lowest
eigenenergy state in a collective motion with a mass~\cite{Won10,Won20}.  From such a viewpoint, a  QED meson particle is  a quantized collective space-time variation of the quark current field
$j^\mu(x)$ or the 
gauge fields $A^\mu(x)$ in a Wheelerian ``particle without particle'' description. 

\section{Generalizing the Schwinger Confinement Mechanism from Quarks in QED in (1+1)D to (QED+QCD) in (1+1)D}\label{sec3}

Even though QCD is a non-Abelian gauge theory, many features of the
lowest-energy QCD mesons, such as quark confinement, meson states, and
meson production, mimic those of the Schwinger model for the Abelian
gauge theory in (1+1)D as noted early on by Bjorken, Casher, Kogut,
and Susskind~\cite{Bjo73,Cas74}.  Such a generic Abelian  string feature in
hadrons was first recognized even earlier by Nambu~\cite{Nam70,Nam74}
and Goto~\cite{Got71}.  They indicated that in matters of confinement,
quark--antiquark bound states, and hadron production of the lowest-energy states, an Abelian approximation of the non-Abelian QCD theory
is a reasonable concept.  Furthermore, t'Hooft showed that in a system
of quarks interacting in the SU($N_{\rm color}$) gauge interaction
approximated as a U($N_{\rm color}$) interaction in the large $N_{\rm
  color}$ limit, planar Feynman diagrams with quarks at the edges
dominate, and the QCD dynamics in (3+1)D can be well approximated as
an open-string in QCD dynamics in (1+1)D~\cite{tHo74a,tHo74b}.
Numerical lattice calculations for a quark and antiquark system
exhibit a flux tube structure~\cite{Bal05,Ama13,Car13}, and a flux
tube can be idealized as a structureless string in (1+1)D.  Thus, the
idealization of QCD meson in (3+1)D as an QCD open string in (1+1)D is
a reasonable concept.

We wish to adopt here the quasi-Abelian approximation of non-Abelian
QCD to obtain stable lowest-energy collective excitations of the QCD
gauge fields and quark currents as carried out in~\cite{Won10,Won20}.
We note first that because quarks carry three colors, the current
$j^\mu(x)$ and the gauge field $A^\mu$ are each a $3\times 3$ matrix
with nine matrix elements.  They can be naturally separated out into
the color-singlet component with the generator $t^0$,
\begin{eqnarray}
t^0=\frac{1}{\sqrt{6}}
\left ( \begin{matrix}
                1 & 0 & 0 \cr
                0 & 1 & 0 \cr
                0 & 0 & 1
        \end{matrix}   \right ),
\label{eq9}
    \end{eqnarray}  
 and the color-octet components with the eight Gell-Mann generators
 $t^1, t^2, t^3,...,t^8$.  The nine generators satisfy the
 orthogonality condition: 2{\rm tr}\,$(t^i t^{j}) =\delta^{ij} $, with
 $i,j =0,1,...,8$.  The quark current and gauge field in the color
 space can be {represented by} 
\vspace{-6pt}\begin{subequations}
\begin{eqnarray}
j^\mu_i(x)= \sum_{i=0}^8 j^\mu_i (x) t^i,\\ A^\mu_i(x)= \sum_{i=0}^8
A^\mu_i (x) t^i .
\end{eqnarray}
\end{subequations}

A stable collective oscillation of the quark--QED--QCD system is a
localized periodic oscillation of the nine dynamical components
$j^\mu_i(x)$ and the associated gauge field $A^\mu_i(x)$.  Because the
color-singlet current generates only color-singlet QED gauge field,
whereas the color-octet current generates only color-octet QCD gauge
fields, the color-singlet and color-octet currents and gauge fields
give rise to  independent collective oscillations.

For the QCD dynamics of the color-octet quark current and gauge field
in (1+1)D, we envisage that a stable QCD state leads to a periodic
trajectory in the eight-dimensional $t^1,t^2,...,t^8$ color-octet
space.  Excitations with a change of the orientation in the state
vector in the eight-dimensional color generator space will lead to
states that depend on the angular variables in the color generator space, and they represent color
excitations which lie substantially above the lowest-energy QCD states.  
In contrast, the
trajectories of the lowest QCD energy states are expected to consist
of a variation of the amplitudes of the current and gauge fields,
without the variation of the orientation of the trajectory in the
eight-dimensional color-octet space.
Therefore, with the limited goal of calculating the lowest-energy QCD
states, it suffices to consider QCD dynamics of quarks and gauge
fields with the trajectory in an arbitrary and fixed orientation in
the color generator space, and to allow only the amplitude of the
current and gauge field to vary.

By projecting the eight-dimensional  color generator space onto a single
 arbitrary color generator axis and  limiting the  dynamics along
   that direction, we attain the
   quasi-Abelian approximation which  is a
   reasonable approximation for the lowest QCD states that do
   not involve gluon excitations.  The Nambu--Goto
   string~\cite{Nam70,Nam74,Got71} is an Abelian string, and it provides an adequate description of 
    the lowest QCD states.
    Casher, Kogut
   and Susskind~\cite{Cas74} used the Schwinger Abelian open string as
   a good model for particle production in the fragmentation of a
   quark--antiquark pair.  The Yo-Yo model of hadrons~\cite{Art74} in
   the Lund Model~\cite{And83} used the Abelian strings to describe
   hadron bound states.  Phenomenological non-relativistic and
   relativistic  quark models for the investigation of the low-lying spectrum
   can remain useful by limiting themselves to dynamics without the
   color excitation in the quasi-Abelian approximation.

Among the eight generators in the color-octet generator space, $t^1$ is
just as randomly and arbitrarily oriented as any other color-octet
generator.  The generator $t^1$ can be taken as a generic generator
for the color-octet sector.  We therefore represent the QCD current
and gauge fields by the $j_1^\mu (x) t^1$ and $A_1^\mu (x)t^1$ along
only the $t^1$ direction.  Because $t^1$ commutes with itself, the
dynamics of the current and gauge fields in QCD with the $t^1$
generator is Abelian.  Fixing the orientation of the generator vector
$t^1$ as unchanged is a quasi-Abelian approximation of the non-Abelian
QCD dynamics in a subspace, in which the dynamics of the quark-QCD
systems can lead to stable QCD collective excitations of the lowest-energy~states.

Upon including (i) the above
QCD current $j^\mu_1(x)$ and gauge field $ A^\mu_1$ associated with
the color-octet SU(3) generator $t^1$ to describe the quasi-Abelian  QCD dynamics as
in~\cite{Won10,Won20}, and also (ii) the QED current $j^\mu_0(x)$ and gauge field $
A^\mu_0$ associated with the color-singlet U(1) generator $t^0$ to
describe the QED dynamics of the color-singlet component, 
we have the sum current and gauge field
\vspace{-3pt}\begin{subequations}
\begin{eqnarray}
j^\mu (x) =  j^\mu_0 (x) t^0 + j^\mu _1 (x)t^1,
\\
A^\mu (x) =  A^\mu_0 (x) t^0 + A^\mu _1 (x)t^1,
\label{2v}
\end{eqnarray}
\end{subequations}
where the generators $t^0$ and $t^1$ satisfy 2{\rm tr}$(t^\lambda
t^{\lambda'}) =\delta^{\lambda {\lambda'}} $, with
$\lambda,{\lambda'}=0,1$.  Subject to the above applied disturbing
gauge field $A^\mu(x)$ with both the QED and the QCD interactions
within the quasi-Abelian approximation in (1+1)D, the massless quark
field $\psi(x)$ satisfies the Dirac equation,
\vspace{-12pt}\begin{eqnarray}
\gamma_\mu ( p^\mu - \sum_{\lambda=0} ^1g_{\2d}^\lambda  A_\lambda ^\mu t^\lambda) \psi =0,
\label{eq77}
\end{eqnarray}
where the quark field $\psi$ is a column vector in color space, and
$\lambda=0$ for the QED interaction, and $\lambda=1$ for the QCD
interaction.  The massless quark field leads to the gauge-invariant
induced quark current $ j^\mu_\lambda (x)=2 {\rm tr} \langle \bar \psi
(x) \gamma^\mu t^\lambda \psi (x) \rangle $ which can be evaluated and found
to relate explicitly to the applied QED gauge field $A^\mu(x)$ by
\cite{Won10,Won20,Won23}
\begin{eqnarray}
j^\mu_\lambda (x)  = -\frac{g_\2d^\lambda}{\pi }\left ( A^\mu_\lambda  (x) - \partial ^\mu \frac{1}{\partial _\eta \partial ^\eta} \partial _\nu A ^\nu_\lambda (x) \right ),~~~ \lambda=0 ~{\rm for ~QED}, {\rm and}~ \lambda=1  ~{\rm for ~QCD}.
\label{eqxx}
\end{eqnarray}

The induced current $j^\mu_\lambda$ generates a gauge field $\tilde
A^\mu_\lambda (x) $ through the Maxwell equation,
\begin{eqnarray}
\hspace*{-0.3cm}\partial_\nu F^{\nu \mu}_\lambda (x) =\partial_\nu  \{ \partial^\nu \tilde A^\mu_\lambda (x)  -  \partial^\mu \tilde A^\nu_\lambda (x) \}=g_\2d^\lambda j_\lambda^\mu (x) .
\label{eqyy}
\end{eqnarray}

A stable collective particle--hole excitation of the quark system
occurs when the initial applied gauge field $A_\lambda^\mu(x)$ gives
rise to the induced quark current $j_\lambda^\mu$, which in turn leads
to the new gauge field $\tilde A_\lambda^\mu(x)$ self-consistently.
We impose this self-consistency condition for the gauge field,
$A_\lambda^\mu(x)=\tilde A_\lambda^\mu(x) $.  In that case,
Equations~(\ref{eqxx}) and (\ref{eqyy}) are the same as
\begin{eqnarray}
&&\partial _\nu \partial ^\nu A_\lambda ^\mu (x) + \frac{(g_\2d^\lambda)^2}{\pi }A_\lambda^\mu(x) =0, 
\\
\hspace*{-4.5cm}{\rm and} ~~~~~~~~~~~~~~~~~~~~~~~~~~~~~~~
&&\partial _\nu \partial ^\nu j_\lambda^\mu (x) +\frac{(g_\2d^\lambda)^2}{\pi }j_\lambda^\mu (x) =0,\!
\end{eqnarray}
where both $j_\lambda^\mu(x)$ and $A_\lambda^\mu(x)$ satisfy the
Klein--Gordon equation for a bound and confined boson with a mass
$m^\lambda =g_\2d^\lambda/\sqrt{\pi}$ given by
\vspace{-6pt}\begin{eqnarray}
m^\lambda  = \frac{g_\2d^\lambda}{\sqrt{\pi}}.
\end{eqnarray}

From the above analysis, we find that the QED and the QCD interactions
separate out as independent interactions, and they contain independent
collective oscillations of the color-singlet current and the
color-octet current, with different oscillation energies proportional
to the different coupling constants of the interactions.  Such a
separation is possible because the relation between $j^\mu$ and
$A^\mu$ as given by Equations~(\ref{eqxx}) and (\ref{eqyy}) are linear
in form but the differential operator in the second term of
Equation~(\ref{eqxx}) contains non-trivial and non-linear differential
elements, which lead to a self-confined local current that can execute
stable collective dynamics of QED and QCD mesons in (1+1)D.

\section{Do the QED and QCD Mesons in (1+1)D Represent  Physical Mesons in (3+1)D?}\label{sec4}

From the results in the last section, we know that there can be stable
collective QED and QCD meson states in (1+1)D whose masses depend on
the coupling constants.  They represent different excitations of the
quark--QED--QCD medium and can be alternatively depicted as open-string
QED and QCD states in (1+1)D, with the quark and the antiquark at
the two ends of the string.

Can these open-string meson states in (1+1)D be the idealization of
physical QCD and QED mesons in (3+1)D?  An open-string QCD meson in
(1+1)D as the idealization of a physical QCD meson in (3+1)D is a
generally accepted concept, as discussed in the last section.  The
observations of the anomalous soft photons, the hypothetical X17
particle, and the hypothetical E38 particle suggest it useful to study
the possibility that the QED mesons in (1+1)D may represent
physical mesons in the physical world of (3+1)D, just as the QCD~mesons.

We shall therefore consider the phenomenological open-string model of
QCD and QED mesons in (1+1)D and study whether they may represent
physical mesons in (3+1)D by comparing theoretical predictions of the
model with experiments.

We note that in (3+1)D, the flux tube has a structure with a
transverse radius $R_T$ and the coupling constant $g_\4d$ is
dimensionless.  In (1+1)D, however, the open string has no structure,
but the coupling constant $g_\2d$ acquires the dimension of a mass.
In an earlier work~\cite{Won09}, when we compactify a system
with a flux tube in (3+1)D with cylindrical symmetry into a (1+1)D
system of a string without a structure, we find that the Dirac equation of
for quarks in (3+1)D can be separated into the coupling of the
longitudinal and the transverse degrees of freedom.  Upon quantizing
the transverse degree of freedom to obtain the lowest 
transverse eigenstates, the longitudinal Dirac equation in (1+1)D
contains an effective transverse mass, and a modified coupling
constant $g_\2d$ which depends on the coupling constant $g_\4d$ in
(3+1)D multiplied by the absolute square of the transverse wave
function.  As a consequence, the longitudinal equation in (1+1)D
contains a coupling constant in (1+1)D that encodes the information of
the flux tube transverse radius $R_T$ and the coupling constant in
(3+1)D in Equation~(34) of~\cite{Won09}  as
\begin{eqnarray}
(g_{\2d})^2=\frac{1}{\pi
    R_T^2}(g_{\4d})^2=\frac{4\alpha_{\4d}}{R_T^2},
\label{12}
\end{eqnarray}
whose qualitative consistency can be checked by dimensional analysis
and by subsequent theoretical studies in  Appendix B of~\cite{Won23}.  
The compactification   from four-dimensional  space-time with cylindrical symmetry 
to two-dimensional with supplementary transverse degrees of freedom has also been examined 
from  the action principle viewpoint, and the relation Equation~(\ref{12})  relating the coupling constant 
for the longitudinal system in 2D with the coupling in 4D has also been found valid from the viewpoint of the action  integral~\cite{Kos12,Won23}.

We can give here a simplified derivation of the above Equation~(\ref{12}).  We consider only the $(x^0,x^3)$ degrees of freedom after the separation of the transverse degree of freedom, and we take the Coulomb gauge, $A^3=0$.  The Maxwell equation for the gauge field 
$A_\4d^0$  in (3+1)D involves  the 4D coupling constant $g_\4d$ and the density $j_\4d^0$  given by   
\begin{eqnarray}
\partial_3^2 A_\4d^0 = g_\4d j_\4d^0.
\label{eq1}
\end{eqnarray}   

In the compactified (1+1)D string picture, there is a similar Maxwell equation  involving
$ A_\2d^0$, the 2D coupling constant $g_\2d$, and the density $j_\2d^0$ , given by
\begin{eqnarray}
\partial_3^2 A_\2d^0 = g_\2d j_\2d^0.
\label{eq2}
\end{eqnarray}  

The (3+1)D charge density  $j_\4d^0$ and  the  (1+1)D charge density  $j_\2d^0$ are related by
\vspace{-6pt}
\begin{eqnarray}
 j_\4d^0 = \frac{1}{\pi R_{T}^2} j_\2d^0.
 \label{eq3}
\end{eqnarray}

When we compactify the Dirac equation for the quark from (3+1)D to (1+1)D, 
we identify 
the gauge interaction term $g_\4d A_\4d^0$   in (3+1)D  as $g_\2d A_\2d^0$ 
in (1+1)D~\cite{Won09,Won23}.   That is, we have 
\vspace{-6pt}\begin{eqnarray}
g_\4d A_\4d^0=g_\2d A_\2d^0.
\label{eq24}
\end{eqnarray}

The consistent solution of the above four {Equations} 
 (\ref{eq1})--(\ref{eq24}) leads to the relation between the coupling constants $g_\2d$ in (1+1)D and $g_\4d$ in (3+1)D in  Equation~(\ref{12}).

With a flux tube in (3+1)D idealized as a string in (1+1)D, the string
in (1+1)D can be decoded back to the original physical flux tube in
(3+1)D by using the above Equation~(\ref{12}).  The boson mass $m$
determined in (1+1)D is the physical mass related to the physical
coupling constant $\alpha_{\4d}$=$(g_{\4d})^2/4\pi$ and the flux tube
radius $R_T$ in (3+1)D by~\cite{Won10}
\begin{eqnarray}
m^2= \frac{4\alpha_{\4d} }{\pi R_T^2}.
\end{eqnarray}

With $\alpha_{\4d}^{\qed}$\!=$\alpha_{{}_{\rm QED}}$=1/137,
$\alpha_{\4d}^{\qcd}$\!=$\alpha_s $$\sim$0.6 from hadron spectroscopy
\cite{Won00,Won01,Cra09,Bal08,Deu16}, and \linebreak  $R_T$$\sim$0.4 fm from
lattice QCD calculations~\cite{Cos17} and $\langle p_T^2 \rangle $ of
produced hadrons in high-energy $e^+e^-$ annihilations~\cite{Pet88},
we estimate the masses of the open-string QCD and QCD mesons to be
\begin{eqnarray}
m^{\qcd}\sim 431{\rm ~ MeV}, ~~~ {\rm and}~~ m^{\qed}\sim 47 {\rm
  ~MeV}.
\label{14}
\end{eqnarray}

The above mass scales provide an encouraging guide for the present
task of a quantitative description of the QCD and QED mesons as open
strings, using QCD and QED gauge field theories in (1+ 1)D.

\section{Phenomenological Open-String Model of QCD and QED Mesons }

For a more quantitative comparison with experimental QCD and QED meson
masses, we need an open-string model of QED and QCD mesons with many
flavors,  and the dependencies of  quark attributes on   flavors and interactions. 
 By the method of bonsonization, local charge-zero bilinear
operator in the Dirac theory corresponds to some local function in the
boson theory.  The time-like component of the current $j^0$ for quarks
with $N_f$ flavors is~\cite{Col76,Won20,Nag09}
\vspace{-6pt}\begin{eqnarray}
j^0 = \sum_f^{N_f}   Q_f  :\! \bar \psi_f \gamma^0 \psi_f \!: \, =  \frac{ \epsilon^{\mu \nu}} { \sqrt{\pi}}
\sum_f^{N_f}   Q_f \,\partial _\nu\phi_f,
\end{eqnarray}
where $\phi_f$ is the $|q_f \bar q_f\rangle $ boson state with the $f$
flavor, $Q_f$ is the charge number for the quark with the $f$ flavor,
$Q_u^\qcd=Q_d^\qcd=Q_s^\qcd$ for QCD, and $Q_u^\qed=\frac{2}{3}, $ and
$Q_d^\qed=-\frac{1}{3}$ for QED.  The interaction Hamiltonian for the
case of massless quarks with $N_f$ flavors is~\cite{Col76,Won20,Nag09}
\vspace{-6pt}\begin{eqnarray}
~~~~~~~~~{\cal H}_{\rm int}&=& \frac{(g_\2d)^2 }{2}  \int dx ~dy j^0(x) j^0(y) |x-y|
\nonumber\\
&=& \frac{(g_\2d)^2 }{2\pi} (\sum_{f=1}^{N_f} Q_f  \phi_f)^2.
\end{eqnarray}

The physical meson state $\Phi_i$  is  a flavor mixture $D_{ij}$  of the flavor  states $\phi_f$,
\vspace{-6pt}\begin{eqnarray}
\Phi_i = \sum_{f=1}^{N_f} D_{if}\phi_f,   ~~~~~  i=0, ..., i_{\rm max},~~{\rm and}~~ i_{\rm max}=N_f.
\end{eqnarray}

The inverse transformation is
\begin{eqnarray}
\phi_f=\sum_{i=0}^{i_{\rm max} }(D^{-1})_{fi} \Phi_i=\sum_{i=0}^{i_{\rm max}} D_{if} \Phi_i.
\end{eqnarray}

For massless quarks, the interaction Hamiltonian written in terms of
the physical states $\Phi_i$ is
\vspace{-6pt}\begin{eqnarray}
{\cal H}_{\rm int}= \frac{(g_\2d)^2 }{2\pi} (\sum_{f=1}^{N_f} Q_f  \sum_{i=0}^{i_{\rm max}} D_{if} \Phi_i)^2.
\end{eqnarray}

The boson mass square of the physical meson  $\Phi_i$ is, therefore,
\begin{eqnarray}
m_i ^2=\frac {\partial^2 {\cal  H}_{\rm int} }{\partial \Phi_i^{~2}}
=\frac{(g_\2d)^2 }{\pi} (\sum_{f=1}^{N_f} D_{if}Q_f)^2=\frac{(g_\2d)^2 }{\pi} (\tilde Q_{i,{\rm eff}})^2,
\end{eqnarray}
where the effective charge  $\tilde Q_{i,{\rm eff}}$ for the physical state $i$ is given by 
\begin{eqnarray}
\tilde Q_{i,{\rm eff}}= |\sum_{f=1}^{N_f} D_{if}Q_f|.
\label{27}
\end{eqnarray}

In the massless quark limit, the meson mass for the $\lambda$th
interaction (with $\lambda=0$ for QED and $\lambda=1$ for QCD) is
\vspace{-6pt}\begin{eqnarray}
(m_{i}^\lambda)^2=\frac{(g_{\2d}^\lambda)^2}{\pi} \left [
    \sum_f^{N_f}  D_{if}^\lambda Q_{f}^\lambda \right ] ^2 .
\end{eqnarray}

For light quarks with two flavors and isospin symmetry, the physical
isoscalar state $\Phi_0$ with $(I=0,I_3=0)$ and the physical isovector
state $\Phi_1$ with $(I=1,I_3=0)$ are given in terms of $|u\bar
u\rangle$ state $\phi_1$ and $|d\bar d\rangle$ state $\phi_2$ by
\begin{subequations}
\begin{eqnarray}
\Phi_0 = \frac{1}{\sqrt{2}} (\phi_1+\phi_2),\\
\Phi_1= \frac{1}{\sqrt{2}} (\phi_1-\phi_2),
\end{eqnarray}
\unskip
\end{subequations}
\unskip
\begin{eqnarray}
\hfil\hspace*{-0.9cm} {\rm and} ~~~~~~~~~~~
m_i^\lambda&=&
\frac{g_{\2d}^\lambda}{\sqrt{\pi}} \left |
    \sum_f^{N_f}  D_{if}^\lambda Q_{f}^\lambda \right |=
\frac{g_\2d^\lambda}{\sqrt{\pi}}\frac{ |Q_u^\lambda-(-1)^i Q_d^\lambda|}{\sqrt{2}},~~~{\rm for~isospin }~~i=0, 1.
\label{29}
\end{eqnarray}

For the QCD isovector pion state $\Phi_1$, the effective color charge is
$Q_{1,{\rm eff} }^\qcd $=$|\sum_{f=1}^2 (D_{1f} Q_f)|$
=$|1/\sqrt{2}-1/\sqrt{2}|$=0.  
In the massless quark limit, the above mass formula (\ref{29}) gives a
zero pion mass in Table 1, which is consistent with the common perception in QCD
that $\pi^0$ is a Goldstone boson.  {The mass of
$\pi^0$ comes only from the quark mass and chiral condensate}
~which contribute $\sum_{f=1}^2 m_f\langle \bar \psi_f \psi_f \rangle $ to
the Lagrangian density and to the $\pi^0$ mass square
\cite{Won20,Won23}
\vspace{-6pt} \begin{eqnarray}
m_\pi^2 &=&
\sum_{f=1}^2  m_f (D^\qcdu_ {If})^2 \langle \bar \psi _f\psi_f \rangle_{_{\rm QCD}},
\nonumber\\
&=&  m_{ud}\langle \bar \psi \psi \rangle_{_{\rm QCD}},
\label{eq111}
\end{eqnarray}
where 
$m_{ud}$=$(m_u+m_d)/2$ and 
$\langle \bar \psi \psi \rangle_{_{\rm  QCD}}$ is the chiral condensate.  

The above equation is consistent with  the Gell--Mann--Oakes--Renner relation
\cite{Gel68},
\begin{eqnarray}
m_\pi^2= (m_u + m_d) \frac{ |\langle 0|\bar q q |0 \rangle|}{F_\pi^2},
\label{233}
\end{eqnarray}
where $|\langle0| \bar q q |0\rangle|$ is the light $u$ and $d$
quark--antiquark condensate, and $F_\pi$ is the pion decay constant.  We
can use the pion mass $m_\pi$ to calibrate the chiral condensate
$\langle \bar \psi \psi \rangle_{_{\rm QCD}}$.  Therefore, the masses
of neutral QCD mesons are given by
\begin{eqnarray}
m_{\lambda I}^2
=\frac{4\alpha_\lambda }{ \pi R_T^2}  \biggl [\sum_f D^\lambda_{If} Q^\lambda_f \biggr ]^2\,
+m_\pi^2  \frac{\sum_f  m_f (D^\lambda_ {If})^2}{m_{ud}},
\label{112}
\end{eqnarray}

For QCD mesons, Equation (\ref{14}) indicates that the mass scale
$m^{\qcd}\!\!\!\!\sim$ 431 MeV $ \gg m_u, m_d, m_s$.  It is necessary
to include $u$, $d$, and $s$ quarks with $N_f $=3 in the analysis of
open-string QCD mesons.  We denote $\phi_1=|u \bar u \rangle$,
$\phi_2=|d \bar d \rangle$, and $\phi_3=|s \bar s \rangle$, and assume
the standard quark model description of $|\pi^0\rangle$,
$|\eta\rangle$, and $|\eta'\rangle$ in terms of flavor octet and
flavor singlet states, with the mixing of the $|\eta\rangle$ and
$|\eta'\rangle$ represented by a mixing angle $\theta_P$~\cite{PDG19}.
The physical states of $|\pi^0\rangle$, $|\eta\rangle$, and
$|\eta'\rangle$ can be represented in terms of the flavor states
$\phi_1$, $\phi_2$ and $\phi_3$ by
\vspace{-6pt}\begin{subequations}
\begin{eqnarray}
|\pi^0\rangle&=&\Phi_1=\frac{\phi_1-\phi_2}{\sqrt{2}},
\\ |\eta~\rangle&=&\Phi_2= |\eta_8\rangle \cos \theta_P- |
\eta_0\rangle \sin \theta _P,
\label{229b}
\\ |\eta'\rangle&=&\Phi_3= |\eta_8 \rangle \sin \theta_P+|
\eta_0\rangle \cos \theta _P,
\label{229c}
\end{eqnarray}
\unskip
\begin{eqnarray}
\hspace*{-4.0cm}\text{where }\hspace*{4.0cm}
&&|\eta_8\rangle
=\frac{\phi_1+\phi_2-2\phi_3}{\sqrt{6}}, \\ &&\,|\eta_0\rangle
=\frac{\sqrt{2}(\phi_1+\phi_2+\phi_3) }{\sqrt{6}}.
\end{eqnarray}
\end{subequations}

The QCD states $\Phi_i$=$\sum_f D_{if}^\qcd\phi_f$ and the flavor
component states $\phi_f$ are then related~by
\begin{eqnarray}
\hspace*{-0.0cm}
\begin{pmatrix}
 \Phi_1\\ \Phi_2\\ \Phi_3
\end{pmatrix}
\!\!=\!\!\begin{pmatrix} \frac{1}{\sqrt{2} } & - \frac{1}{\sqrt{2} } &
0 \\ \frac{1} {\sqrt{6}} \{ \cos \theta_P \!-\!\sqrt{2} ~ \sin
\theta_P\} & \frac{1} {\sqrt{6}} \{ \cos \theta_P \!-\!\sqrt{2} ~ \sin
\theta_P\} & \frac{1} {\sqrt{6}}\{-2\cos \theta_P \!-\! \sqrt{2}\sin
\theta_P\} \\ \frac{1}{\sqrt{6}}\{\sin \theta_P \!+\! \sqrt{2}\cos
\theta_P\} & \frac{1}{\sqrt{6}}\{\sin \theta_P \!+\! \sqrt{2}\cos
\theta_P\} & \frac{1}{\sqrt{6}}\{-2\sin \theta_P\! +\! \sqrt{2}\cos
\theta_P\} \\
\end{pmatrix}
\!\!
\begin{pmatrix}
 \phi_1\\ \phi_2\\ \phi_3
\end{pmatrix}\!\!,
\nonumber
\end{eqnarray}
with the inverse relation $\phi_f= \sum_{i=1}^3 D_{if}^\qcd\Phi_i$,
\begin{eqnarray}
\hspace*{0.0cm}
\begin{pmatrix}
 \phi_1\\ \phi_2\\ \phi_3
\end{pmatrix}=
\begin{pmatrix}
 \frac{1}{\sqrt{2} } & ~~~ \frac{1} {\sqrt{6}} \{ \cos \theta_P
 -\sqrt{2} ~ \sin \theta_P\}~~~ & \frac{1} {\sqrt{6}}\{ \sin \theta _P
 + \sqrt{2} ~\cos \theta _P \} \\ - \frac{1}{\sqrt{2} } & \frac{1}
 {\sqrt{6}}\{ \cos \theta_P - \sqrt{2} ~\sin \theta_P\} & \frac{1}
 {\sqrt{6}}\{ \sin \theta _P + \sqrt{2} ~\cos \theta _P \} \\ 0
 &\frac{1} {\sqrt{6}}\{-2\cos \theta_P \!-\! \sqrt{2}\sin \theta_P\} &
 \frac{1}{\sqrt{6}}\{-2\sin \theta_P\! +\! \sqrt{2}\cos \theta_P\} \\
\end{pmatrix}
\begin{pmatrix}
 \Phi_1\\ \Phi_2\\ \Phi_3
\end{pmatrix}.~
\end{eqnarray}

For these QCD mesons, there is a wealth of information on the matrix
$D_{if}^\qcd$ that describes the composition of the physical states in
terms of the flavor components as represented by the mixing angle
$\theta_P$ between the flavor octet and flavor singlet components of
the SU(3) wave functions in $\eta$ and $\eta'$ in (\ref{229b}) and
(\ref{229c}). The ratio of the strange quark mass to the light $u$ and
$d$ quark masses that is needed in the above mass formula is also
known.  From the tabulation in PDG~\cite{PDG19}, we find
$\theta_P=-24.5^{\circ}$ and ${m_s}/m_{ud}$= 27.3$_{-1.3}^{+0.7}$.  The only
free parameters left in the mass formula (\ref{112}) are the strong
interaction coupling constant $\alpha_s=\alpha_\4d^\qcd$ and the flux
tube radius $R_T$.

For the value of $\alpha_s$, previous works on the non-perturbative
potential models use a value of $\alpha_s$ of the order of $0.4-0.6$
in hadron spectroscopy studies
\cite{Bar92,Won00,Won01,Cra09,Bal08,Deu16}.  We find that
$\alpha_s=0.68$ gives a reasonable description of the QCD mesons
masses considered.  For the value of $R_T$, lattice gauge calculations
with dynamical fermions give a flux tube root-mean-square radius
$R_T$ = 0.411 fm for a quark--antiquark separation of 0.95 fm
\cite{Cos17}.  The experimental value of $ \langle p_T^2\rangle $ of
produced hadrons ranges from 0.2 to 0.3 GeV$^2$ for $e^+$$-$$e^-$
annihilations at $\sqrt{s}$ from 29 GeV to 90 GeV~\cite{Pet88},
corresponding to a flux tube radius $R_T$ = $\hbar/\sqrt{\langle
  p_T^2\rangle }$ of 0.36 to 0.44 fm.  It is reasonable to consider
the flux tube radius parameter to be $R_T$=0.4 fm.  This set of parameters
of $\alpha_s$ = 0.68 and $R_T$ = 0.40 fm gives an adequate description of
the $\pi^0$, $\eta$ and $\eta'$ masses as shown in the last column of
Table~\ref{tb1}.

\begin{table}[h]
\centering
\caption { Comparison of experimental and theoretical masses of
  neutral, $I_3$=0 QCD and QED mesons obtained in the massless quark
  limit and with the semi-empirical mass formula (\ref{112}) for QCD
  mesons and (\ref{1111}) for QED mesons, for $\alpha_\4d^{{}_{\rm
      QED}}$=1/137, $\alpha_s=\alpha_\4d^\qcd$=0.68, and $R_T$=0.40
  fm.  }
\vspace*{0.2cm}
\begin{tabular}{|c|c|c|c|c|c|c|c|}
\cline{3-6} \multicolumn{2}{c|}{}&  &Experimental
&Meson mass& Meson mass  \\
 \multicolumn{2}{c|}{}&I & mass  &in massless & including quark mass \\
 \multicolumn{2}{c|}{}&  &   & quark limit & \& quark condensate    \\
 \multicolumn{2}{c|}{}& & (MeV) & (MeV) &  contributions (MeV) \\ 
\hline
QCD&$\pi^0$ & 1 &\!\!134.9 \!\! & 0 & 134.9$^\ddagger$\\ 
\!\!meson\!\!& $\eta$  & 0 &\!\!547.9\!\! & 329.7 &498.4~~\\ 
& $\eta'$ &  0  & 957.8 &723.4& 948.2 \\ \hline
QED&\!\!isoscalar\!\!& 0  & & 11.2 & 17.9\\ 
\!\!meson\!\!&\!\!isovector\!\!& 1  & &33.6 & 36.4 \\ 
\hline
Possible& X17 &   &\!\!16.70$\pm$0.35$\pm$0.5$^\dagger$~~\!\! & & \\ 
QED& X17 & &\!\!16.84$\pm$0.16$\pm$0.20$^\#$\!\! & & \\ 
meson& X17 & &\!\!17.03$\pm$0.11$\pm$0.20$^\uplus$\!\! & & \\ 
\!\!candidates& X17 & &\!\!16.7$\pm$0.47$^\oplus$\!\! & & \\ 
             & X17 &  &17.1$\pm$0.7$^\bigtriangledown$~~~& & \\ 
             & E38  &               &  37.38$\pm$0.71$^\bigtriangleup$ & &\\ 
             & E38  &                &  40.89$\pm$0.91$^\ominus$ & &\\ 
             & E38 &                &  39.71$\pm$0.71$^\otimes$ & & \\ \hline
\end{tabular}
\vspace*{0.1cm}

\hspace*{-6.45cm}$^\ddagger$ Calibration mass~~~~~~~~~~~~~~~~~~~~~~~\\
$^\dagger$\,A. Krasznahorkay $et~al.$, Phys.Rev.Lett.116,042501(2016), $^8$Be$^*$ decay\cite{Kra16}\\
\hspace*{-2.19cm}$^\#$A. Krasznahorkay $et~al.$, arxiv:1910.10459, $^4$He$^*$ decay\cite{Kra19}~~~~\\
\hspace*{-2.19cm}$^\uplus$A. Krasznahorkay $et~al.$, arxiv:2209.10795, $^{12}$C$^*$ decay\cite{Kra22}~~~~\\
\hspace*{-1.86cm}$^\oplus$ T. A. Tran$~ et~al.$, ISMD Proceedings (2023), $^{8}$Be$^*$ decay\cite{Tra23}~~~~\\
\hspace*{-0.80cm}$^\bigtriangledown$\,K. Abraamyan~$ et~al.$, arxiv:2311.18632(2023), $p$C,$d$C,$d$Cu$\to$$\gamma \gamma X$\cite{Abr23}~\\ 
\hspace*{-0.70cm}$^\bigtriangleup$\,K. Abraamyan $et~al.$, EPJ Web Conf       204,08004(2019),$d$Cu$\to$$\gamma \gamma X$\cite{Abr19}~\\
\hspace*{-0.70cm}$^\ominus$\,K. Abraamyan $et~al.$, EPJ Web Conf       204,08004(2019),$p$Cu$\to$$\gamma \gamma X$\cite{Abr19}~\\
\hspace*{-0.70cm}$^\otimes$\,K. Abraamyan $et~al.$, EPJ Web Conf       204,08004(2019),~$d$C$\to$$\gamma \gamma X$\cite{Abr19}~\\
\label{tb1}
\end{table}

Having provided an adequate description of the neutral QCD meson
masses, we wish to extrapolate to the unknown sector of the QED
mesons.  We do not know the flux tube radius for the QED mesons.  We
shall proceed by presuming that the flux tube radius may be an
intrinsic property of quarks pending modifications by future
experimental data.  The chiral condensate depends on the interaction
type $\lambda$ and the coupling constant $g_\4d^\lambda$.  We note
that the chiral current anomaly in the chiral current depends on the
coupling constant square, $e^2=g_\4d^2 $ as given in Equation~(19.108) of
ref.~\cite{Pes95}
\vspace{-6pt}\begin{eqnarray}
\partial _\mu j^{\mu 5 3}=-\frac{e^2}{32\pi} \epsilon^{\alpha \beta \gamma\delta } F_{\alpha \beta} F_{\gamma \delta},
\end{eqnarray}
which shows that the degree of non-conservation of the chiral current is
proportional to $e^2$.  It is therefore reasonable to infer that the
chiral condensate term scales as the coupling constant square as
$g_\4d^2$ or $\alpha_\4d^\lambda$, just as the first term.  Hence, we
have~\cite{Won20}
\vspace{-6pt}\begin{eqnarray}
m_{\lambda I}^2= 
\frac{4\alpha_\4d^{{}_{\rm \{QCD,QED\}}}}{ \pi R_T^2}
\left [\sum_{f=1}^{N_f} 
D^\lambda _{If} Q^\lambda_f  \right]^2 
+ m_\pi^2\frac{\alpha_\4d^{{}_{\rm \{QCD,QED\}}}}{\alpha_\4d^\qcdd} 
 \frac{\sum_f^{N_f}  m_f (D^\lambda_ {If})^2}{m_{ud}}.
\label{1111}
\end{eqnarray}

By extrapolating to the QED mesons with $\alpha_\4d^\qedd$=1/137, with
$Q_u$=$2/3$ and $Q_d$=$-1/3$, we find an open-string isoscalar
$I(J^\pi)$=$0(0^-)$ QED meson state at 17.9 MeV and an isovector
$(I(J^\pi)$=$1(0^-), I_3=0)$ QED meson state at 36.4 MeV.  The
predicted masses of the isoscalar and isovector QED mesons in Table 1
are close to the mass of the hypothetical X17 and E38 particles.
The open-string
description of the QCD and QED mesons may be  a reasonable concept and the
anomalous X17~\cite{Kra16} and E38~\cite{Abr19} observed recently may
be QED mesons.  The parent particles of the anomalous soft photons
\cite{DEL10} may be QED mesons.

   The flux tube radius of the QED meson is a
   phenomenological parameter.  The fact that the same flux tube
   radius used for QCD-confined mesons can describe also the
   QED-confined QED mesons, together with the non-observation of
   fractional charges, may suggest the possibility that the confinement
   property and the flux tube radius may be intrinsic properties of
   the quarks.

It is of interest to note the different ordering of the isoscalar and
isovector mesons in QCD and QED.  For QCD mesons, because the color
charges of $u$ and $d$ quarks are equal, the effective color charge
for the isovector QCD meson is zero and the effective color charge for
the isoscalar QCD meson is large and non-zero.  Consequently, for the
QCD mesons, $m_{I=1}^\qcd <m_{I=0}^\qcd$.  However, for the QED mesons,
because the electric charges of $u$ and $d$ quarks are opposite in
signs and different in magnitudes, the effective electric charge for the isoscalar state is small,
while the effective electric charge for the isovector state is
relatively large, and we have the ordering $m_{I=1}^\qed > m_{I=0}^\qed$, which is the opposite of the ordering of the QCD mesons.

\section{Production,  Decay, and Detection  of the QED Mesons}

To search for QED mesons, it is necessary to consider the production
and the decay of the QED mesons.  As composite particles of quarks and
antiquarks, they are produced when quark and antiquark pairs at the
proper QED meson eigenstate energies are produced.  Therefore, we
expect QED mesons to accompany the production of QCD mesons in
hadron--hadron collisions, $e^+e^-$ annihilations, and the coalescence
of quark and antiquarks at the deconfinement-to-confinement phase
transitions.  Experimentally, anomalous soft photon production indeed
accompanies QCD hadron production.  The transverse momentum
distribution of the anomalous soft photons also suggests the
occurrence of QED mesons in these reactions~\cite{Won20,Dya21}.
However, the anomalous soft photons  provide only indirect evidence
because their masses have not been directly measured, and their decay
properties not  explicitly demonstrated.  Possible direct
evidence for the QED meson may come from the ATOMKI and HUS
observations of the hypothetical X17 {particle}
~\cite{Kra16,Tra23} as
well as the DUBNA observations of anomalous structures at $\sim$17 and
$\sim$38 MeV~\cite{Abr23}.
 
In ATOMKI and HUS experiments, proton beams at a laboratory energy
of 0.5 to 4.0 MeV were used to fuse a proton with $^3$H, $^7$Li, and
$^{11}$B target nuclei to form excited states of $^4$He, $^8$Be, and
$^{12}$C, respectively~\cite{Kra16,Kra19,Nag19,Kra21,Sas22,Kra22,X1722,X17Kra,Kra23,Tra23}.  The product
nuclei are alpha-conjugate nuclei, and the reactions take place at
energies below or just above the Coulomb barrier.  The alpha-conjugate
nuclei have been so chosen that the   compound nuclei produced after proton fusion are
highly excited, whereas the compound nuclei ground states are closed-shell nuclei of
different shapes with extra stability.  In particular, the $^4$He
ground state is a spherical closed-shell nucleus.  The $^8$Be ground
state is a prolate closed-shell nucleus with a longitudinal-to-transverse radius ratio of about 2:1.  The $^{12}$C ground state is an
oblate closed-shell nucleus with a longitudinal-to-transverse radius
ratio of about 1:2~\cite{Won70}. There is consequently a large
single-particle energy gap between particle states above the closed
shell and the hole states at the top of the fermi surface below the
closed shell.  The captured proton in the
proton fusion reaction populates a proton single-particle state above
the large closed-shell energy gap, and the proton hole is located at the
top of the Fermi surface below the closed shell.  The transition of
the captured proton from the proton particle state to reach the proton hole state below the closed-shell fermi
surface will release the large shell-gap energy that
is of the order of about 17-20 MeV.  Such a large shell--energy gap may be
sufficient to produce a neutral boson if there were to exist such a
stable neutral boson particle with the proper energy, quantum numbers,
and other conditions appropriate for its production.  In the
spatial configuration space, the captured proton in the valence orbit
is located an an extended distance from the alpha conjugate hole
nuclear core to facilitate the possible formation of a flux tube
structure between the proton and the nuclear core to favor the
production of a $q\bar q$ open string.

\begin{figure} [h]
  \centering
\includegraphics[scale=0.65]{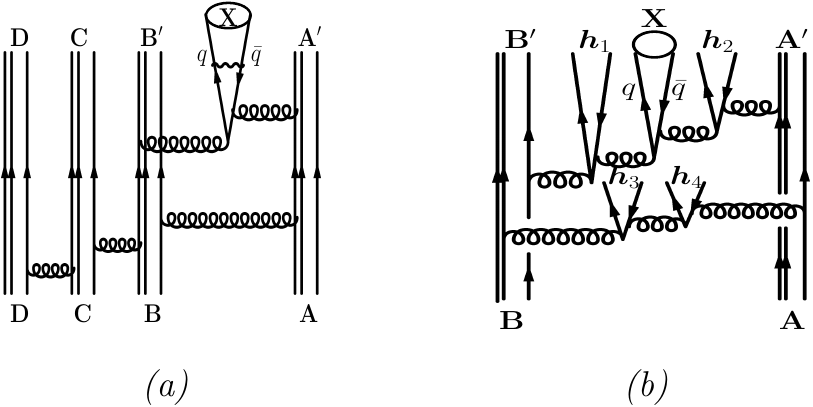}
\caption{(\textbf{a}) $q\bar q(X)$ {production by the fusion} 
 of two virtual
  gluons in the de-excitation of a highly excited $^4$He$^*({\rm
    ABCD}) \to ^4$He (ground state) ({\rm A'B'CD}) + $q$$\bar q$(X) with
  the fusion of two virtual gluons between B and A. (\textbf{b}) $q\bar
  q(X)$ production in hadron--hadron or a nucleus--nucleus
  collision by A + B $\to$ A' + B' + $(q\bar q)^n \to$ A' + B' +$
  \sum\limits_i h_i + q \bar q(X)$. }
\label{fig1}
\end{figure}

As an example of a possible QED meson production process in a
low-energy $p$+$^3$H proton fusion reaction at ATOMKI and HUS, we
show in Figure~\ref{fig1}a the Feynman diagram of the excited
state of $^4$He nucleus, which has been prepared by placing a proton in
the stretched-out $p$ state interacting with the $^3$H core in
Figure~\ref{fig1}a~\cite{Kra16,Kra19,Kra21}.  The de-excitation of
the $^4$He$^*$ excited state to the $^4$He(gs) ground state can occur
by the proton emitting a virtual gluon which fuses with the virtual
gluon from the $^3$H core, leading to the production of a $q\bar q$
pair as shown in Figure~\ref{fig1}a.  If the
$^4$He$^*$$\to$$^4$He(gs) de-excitation energy exceeds the mass of a
confined QED meson $q\bar q$, then a QED meson may be produced in
conjunction with the $^4$He$^*$$\to$$^4$He(gs) de-excitation as
depicted in Figure~\ref{fig1}a.

In the high-energy nucleus--nucleus collision experiments at Dubna~\cite{Abr12,Abr19}, proton and light ion beams collide with light
internal nuclear targets at the JINR Nuclotron at an energy of a few
GeV/nucleon, and many $q\bar q$ pairs are produced as depicted
schematically in~Figure~\ref{fig1}b,
\vspace{-6pt}\begin{eqnarray}
A + B \to A' + B' + (q\bar q)^n .
\end{eqnarray}

The invariant masses of most of the produced $q\bar q$ pairs will
exceed the pion mass, and they will materialize as QCD mesons, 
labeled as $h_i$ in Figure~\ref{fig1}b.  However, there may remain
a small fraction of the color-singlet $[q\bar q]^1$ pairs with an
invariant masses below $m_\pi$.  The $q \bar q$ pairs in this energy
range below mass $m_\pi$ allow the quark and the antiquark to interact
non-perturbatively in QED alone, with the QCD interaction as a spectator interaction, to lead to possible QED
meson eigenstates labeled schematically as $q\bar q(X)$ in~Figure~\ref{fig1}b.

In other circumstances in the deconfinement-to-confinement phase
transition of the quark--gluon plasma in high-energy heavy-ion
collisions, a deconfined quarks and a deconfined antiquark in close
spatial proximity can coalesce to become a $q\bar q$ pair with a pair
energy below the pion mass, and they can interact non-perturbatively in QED alone to lead
to a possible QED meson if there is QED-confined eigenstate in this
energy range.

\begin{figure}[h]
\centering
\includegraphics[scale=0.90]{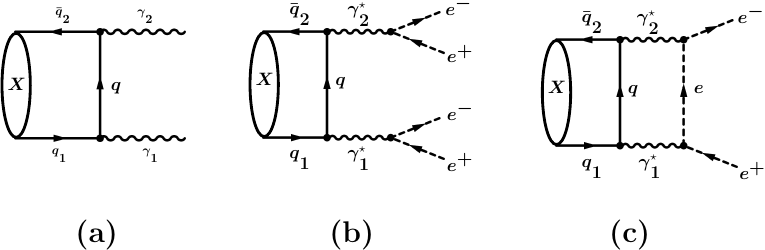}
\caption{(\textbf{a}) A QED meson $X$  can decay into two real
  photons $X \to \gamma_1 + \gamma_2$. (\textbf{b}) It can decay into two
  virtual photons, each of which subsequently decays into a $(e^+ e^-)$ pair,
  $X \to \gamma_1^* + \gamma_2^* \to (e^+ e^-) + ( e^+ e^-) $, and
  (\textbf{c}) it can decay into a single $(e^+ e^-)$ pair, $X \to \gamma_1^*
  + \gamma_2^* \to e^+ e^-$.  }
\label{fig2}
\end{figure}

A QED meson may be detected by its decay products from which its
invariant mass can be measured, even though a QED meson in (1+1)D cannot decay, as the quark
and the antiquark execute yo-yo motion along the string.  In the
physical (3+1)D, the structure of the flux tube and the transverse
photons must be taken into account, in which case the quark and the
antiquark at different transverse coordinates in the flux tube
traveling from opposing longitudinal directions in a QED meson can
make a turn to the transverse direction, by which the quark and the
antiquark can meet and annihilate, leading to the emission of two real
transverse photons $\gamma_1\gamma_2$ as depicted in the Feynman
diagram {Figure}
~\ref{fig2}a.  A QED meson can decay into two virtual
photons $\gamma_1^*\gamma_2^*$, each of which subsequently decays into
an $e^+e^-$ pair as $(e^+e^-)$$(e^+e^-)$ shown in
{Figure}~\ref{fig2}b.  The coupling of the transverse photons to an
electron pair leads further to the decay of the QED meson into an
electron--positron pair $e^+e^-$ as shown in {Figure}~\ref{fig2}c.  The
mass of a QED meson can be obtained by measuring the invariant mass of
its decay products.

\section{Experimental Evidence for the  Possible Existence of the QED Mesons}
\subsection{The ATOMKI Observation of the X17 Particle by $e^+e^-$ Measurements}

Since 2016, the ATOMKI Collaboration has been observing the
occurrence of a neutral boson, the hypothetical X17,  with a mass of about 17 MeV, 
 by studying the $e^+ e^-$ spectrum in the
de-excitation of the excited alpha-conjugate nuclei $^4$He$^*$,
$^8$Be$^*$, $^{12}$C$^*$ at various energies in low-energy
proton fusion experiments~\cite{Kra16,Kra21,Kra22,Kra23}.  A summary
and update of the ATOMKI results has been presented
\cite{X17Kra,Kra23}, and the confirmation of the ATOMKI data for the
$p+{}^7$Be reaction by Hanoi University of Science has been reported
\cite{Tra23}.  Many distribution functions of the opening angle
$\theta_{e^+e^-}$ between $e^+$ and $e^-$ have been successfully
described as the production of the X17 particle with an invariant mass
of about 17 MeV.  The signature for the X17 particle consists of a
resonance structure in the invariant mass of the emitted $e^+e^-$ pair.  Such a signature
is a unique identification of a particle.  Because $e^+e^-$ pairs are
also produced in many other reaction processes, it is important to
subtract contributions from these known processes and from random and
cosmic ray backgrounds.

\begin{figure}[h]
\centering
\includegraphics[scale=0.50]{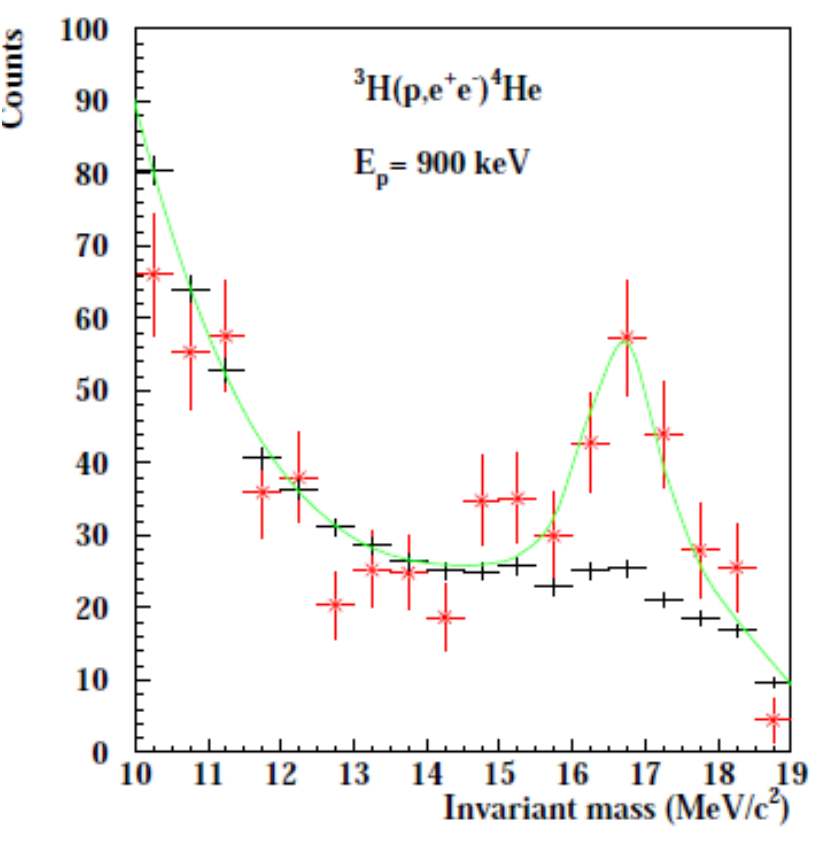} 
\caption{{The invariant mass distribution} 
 of the emitted $e^+$ and
  $e^-$ in the de-excitation of the compound nucleus $^4$He$^*$ state
  at 20.49 MeV in the $^3$H($p,e^+e^-){}^4$He$_{\rm gs}$ reaction at
  $E_p^{\rm lab}=$ 0.9 MeV~\cite{Kra19}.  Red data points are the data
  in the signal region $19.5 < E_{e^+e^-}({\rm sum})<$ 22.0 MeV, and
  black points are data in the background region $5 < E_{e^+e^-}({\rm
    sum})<$ 19.0 MeV .  The solid (green) curve is the fit to the invariant mass data points. }
\label{thetaK}
\end{figure}

For an optimal detection of the X17 signals, the ATOMKI Collaboration
found it necessary to focus on certain regions of the phases space
with strong signals so as to enhance the observation probability.  For
example, in the collision of $p$ with $^3$H at 0.9 MeV
energy, the ATOMKI Collaboration found that the correlation angle
$\theta_{e^+e^-}$ of 120$^{\circ}$ was optimal for a large signal. At such
an angle, the energy sum of the $e^+$ and $e^-$, $E_{e^+e^-}({\rm
  sum})=E_{e^+}+E_{e^-}$ showed a peak structure at around $20.6$ MeV
as shown in Figure~1 of~\cite{Kra21}.  Two spectra were constructed for
the energy sum $E_{e^+e^-}({\rm sum})$, one at
$\theta_{e^+e^-}$=120$^{\circ}$ and another at 60$^{\circ}$, where no X17 signal was
expected.  The energy sum spectrum in the lower panel of in Figure~1 of
ref.~\cite{Kra21} was obtained by subtracting the latter from the former,
after proper normalization.  In the signal region of $19.5 \le
E_{e^+e^-}({\rm sum} )\le 22.0$ MeV and the background in $5 \le
E_{e+e^-}({\rm sum})<19$ MeV, the invariant mass spectrum of the
emitted $e^+$ and $e^-$ showed a resonance structure at $\sim$ 17 MeV
as shown in Figure~\ref{thetaK}.

 Previous analysis on the minimum $e^+e^-$ opening angle in X17 decay  by Feng et~al.~\cite{Fen20}
supports the validity of the ATOMKI X17 emission model~\cite{Kra16}.
Similar analysis of the X17 emission to $e^+e^-$ was also carried out by Barducci and Toni~\cite{Bar23}.
We wish to test here the extended ATOMKI X17 emission model~\cite{Kra23},
which proposes  the emission of the  X17 particle
 not only in the de-excitation of  the produced compound nucleus 
 to the ground state but also to an excited state of the compound nucleus~\cite{Kra16}. 
In the  ATOMKI emission model,  the nature of
the X17 and the coupling between 
the emitting excited alpha conjugate nucleus and 
the emitted X17
particle 
are left unspecified.  Among many possibilities, the X17 in the model
could be the isoscalar QED meson, which has a predicted mass of
$\sim$17~MeV and can decay into $e^+$ and $e^-$~\cite{Won10,Won20}.

In ATOMKI experiments involving the detection of 
 $e^+$ and $e^-$, we envisage that
the reaction first goes through the fusion of
the projectile nucleus $A$  with the target nucleus $B$, producing a compound nucleus
$C^*$, which can be either on a resonance or off a resonance among the
continuum states.  The alpha-conjugate compound nuclei $C^*$ formed by
a low-energy proton fusion in the ATOMKI experiments are expected to be essentially simple shell
model states. 
  For example, the lowest ${}^4$He$^*$ states would have
a large component as a proton in the $p$-shell orbiting a spherical
${}^3$H core, while the lowest-energy $^8$Be$^*$ and $^{12}$C$^*$
states would have large components of an excited $^4$He$^*$ with one
and two spectator alpha clusters, respectively.
However,  more complicated states with other light projectile may also be possible with other light
projectiles and light  targets at different energies.

The dominant de-excitation of the excited compound nucleus
$C^*$ to produce  $e^+$ and $e^-$ is by way of the emission of a photon $\gamma$ to
 de-excite radiatively to the compound
nucleus $C_{\rm f}$ state and the conversion of the photon to $e^+e^-$ by internal pair conversion.  
Here, the $C_{\rm f}$ state can be
an
excited state  or
 the ground state $C_{\rm gs}$ 
of the compound nucleus
$C$.  If the final state $C_{\rm f}$  is an excited state, then there can also be
an additional $\gamma_{\rm f}$ emission, for example, with $E(\gamma_{\rm
  f})=M(C_{\rm f})-M(C_{\rm gs})$.  The emitted photon (or photons)
can be the source of $e^+e^-$ pairs.  By the internal pair conversion
process, the emitted photon can be internally converted into an $e^+
e^-$ pair,
\begin{eqnarray}
 A + B \to C^* \to  C_{\rm f} +\gamma ,~~~ {\rm followed~by} ~~~\left \{ \gamma\xrightarrow{{\rm internal ~conversion}}  e^+ + e^- ~~\text{and}~~  C_{\rm f} \to C_{\rm gs} + \gamma_{\rm f} \right \}.
\nonumber
\end{eqnarray}

The internal pair conversion processes have characteristic $e^+e^-$
angular correlations that depend on the multi-polarity of the
$C^*$$\xrightarrow{\gamma} C_{\rm f}$ transition.  They are relatively
smooth distribution functions of the opening angle $\theta_{e^+e^-}$
between $e^+$ and $e^-$ in the CM system, with no abrupt rises and falls as a function
of the opening angle.

In the presence of the internal pair conversion $e^+e^-$ background,
the ATOMKI Collaboration searches for an unknown neutral boson X that
may be emitted by $C^*$ in its de-excitation to the lower $C_{\rm f} $
final state or the $C_{\rm gs}$ ground state,
\begin{eqnarray}
A + B  \to C^* \to   C_{\rm f} +  { X}  ,  ~~\text{followed by } ~~\left \{  X \to e^+ + e^- ~~\text{and}~~  C_{\rm f} \to C_{\rm gs} + \gamma_{\rm f}\right \}.
\end{eqnarray}

The subsequent decay of the $X$ particle into $e^+$ and $e^-$ will
give the X  emission  signal in the angular distribution of
$\theta_{e^+e^-}$ between $e^+$ and $e^-$ in ATOMKI-type  experiments.

We begin first by studying the X particle decaying into $e^+$ and
$e^-$ in the $X$ rest frame.  We denote the four-momenta of $e^+$ and
$e^-$ by $(\epsilon_{e^+},\vec p_{e^+})$ and $(\epsilon_{e^-},\vec
p_{e^-})$, respectively.  Because $m_X \gg m_e$, we can neglect the
mass of the electron and consider $\epsilon_{e^+}=\epsilon_{e^-}=
\frac{m_X}{2}$ $\approx$ $|\vec p_{e^+}|$ = $|\vec p_{e^-}|$ with $\vec p_{e^+}$ 
pointing in the $(\theta,\phi)$ direction, and $\vec p_{e^-}= -\vec p_{e^+}$.

To obtain the opening angle $\theta_{e^+e^-}$ in the compound nucleus
$C^*$ rest frame, which is also the same as the $A$+$B$ reaction
center-of-mass (CM) frame, we consider next the decay of the compound
nucleus $C^*$ to $C_{\rm f}$ and $X$.  By the decaying process, the
$X$ particle acquires a kinetic energy $K$, and correspondingly, a
three-velocity $\vec \beta$, which we can take to lie along the $z$-axis,
with $\vec \beta=\beta \vec e_z$.

To obtain the four-momenta of $e^+$ and $e^-$ in the CM frame, we boost
the four-momenta of $e^+$ and $e^-$ in the $X$ rest frame by the
three-velocity $\vec \beta$ of the $X$ particle in the CM system.  The
three-momenta $\vec p_{e^{\pm {}}}'$
of the emitted $e^+$ and $e^-$ in the CM system are therefore
\begin{eqnarray}
\vec p_{e^{\pm {}}}'=  \left \{  \vec p_{e^{\pm {}}}+  \vec \beta  \gamma ( \frac{\gamma}{\gamma+1}  \vec \beta \cdot  \vec p_{e^{\pm {}}} + \epsilon_{e^{\pm}} )  \right \}.
\label{eq42}
\end{eqnarray}
where $\gamma=1/\sqrt{1-\beta^2}$.  The
squares of the energies and the momenta of $e^+$ and $e^-$ in the CM
frame are therefore
\vspace{-6pt}
\begin{eqnarray}
\hspace*{0.0cm}
(\epsilon_{e^\pm}')^2\!\!\approx \!\!|\vec p_{e^{\pm {}}}'|^2\!\!\!= \!\!\left ( \frac{m_X}{2}\right )^2  \left \{  1  \pm 2   \beta \cos \theta   \gamma ( \pm \frac{\gamma}{\gamma+1}  \beta \cos \theta  + 1 )  +  \beta^2  \gamma^2 ( \pm \frac{\gamma}{\gamma+1}  \beta \cos \theta  + 1 )^2  \right \} .
\label{eq43}
\end{eqnarray}
where  we have used $\vec \beta \cdot \vec p_{e^+}=\beta  |\vec p_{e^+}| \cos \theta.$
The scalar product of $\vec p_{e^+}'$ and  $\vec p_{e^-}'$ is
\vspace{-9pt}
\begin{eqnarray}
\hspace*{-3.50cm}
\vec p_{e^+}' \cdot  \vec p_{e^-}'=|\vec  p_{e^+}' |~ | \vec p_{e^-}'| \cos \theta_{e^+e^-} 
\nonumber
\end{eqnarray}
\unskip
\begin{eqnarray}
\hspace*{0.0cm}
=
( \frac{m_X}{2})^2  \left \{  \vec e_z +  \vec \beta  \gamma (  \frac{\gamma}{\gamma+1}  \beta \cos \theta  + 1 )  \right \} \cdot  \left \{ -\vec e_z +  \vec \beta  \gamma ( - \frac{\gamma}{\gamma+1}  \beta \cos \theta  + 1 )  \right \}.
\label{eq44}
\end{eqnarray}

The above equations give the relation between the $X$ particle
velocity angle $\theta$ and the $e^+e^-$ opening angle
$\theta_{e^+e^-}$ in the CM frame, from which the distribution of the opening angle in
$\theta_{e^+e^-}$ in the CM frame can be obtained, when the distribution function of
the $X$ particle velocity angle $\theta$ is known.

We can consider the emission of the X particle from the compound
nucleus $C^*$ to be isotropic in the CM frame until experimental data
provide us with information on the degree of non-isotropy.  Even if the X particle may be emitted in a non-zero angular momentum $l$ or when the compound system is captured in an angular momentum $l$ state, the weights of different $l_z$ components may remain  equal
if there is no preferred target nucleus direction, resulting in an isotropic distribution in the CM system.  The isotropic
emission corresponds to a distribution of the polar angle  $\theta$  in the CM frame
as
\vspace{-6pt}
\begin{eqnarray}
\frac{dN}{d\Omega_\theta}=\frac{dN}{d\cos \theta d \phi }=\frac{1}{4\pi},
\end{eqnarray}
from which the angular distribution of the opening angle $\theta_{e^+e^-}$ can be obtained as
\begin{eqnarray}
\frac{dN}{d\Omega_{\theta_{e^+e^-}}}= \frac{1}{4\pi} \frac{d \cos \theta } {d \cos \theta _{e^+e^-}},
\end{eqnarray}
where the right-hand side can be obtained by using Equations (\ref{eq42})--(\ref{eq44}).

We can infer from Equations (\ref{eq43}) and (\ref{eq44}) that the
maximum opening angle in the CM frame between $e^+$ and $e^-$ occurs when $\vec \beta$
is along the \emph{z}-axis.  For such a case, the opening angle between $e^+$
and $e^-$ is
\vspace{-6pt}\begin{eqnarray}
\theta_{e^+e^-}({\rm max}) =\pi.
\end{eqnarray}

Furthermore, from Equations (\ref{eq43}) and (\ref{eq44}), the minimum
opening angle occurs at $\theta=\frac{\pi}{2}$, which happens when
$\vec \beta$ is perpendicular to $\vec p_{e^+}$ with $\cos \theta = 0$.
The minimum opening angle $\theta_{e^+e^-}$(min) in the CM frame is
given by
\begin{eqnarray}
\cos [\theta_{e^+e^-}({\rm min})] = \frac{ -1 + \beta^2 \gamma^2}{1+\beta^2\gamma^2}= -1 + 2\beta^2,
~~~{\rm and}~~~\theta_{e^+e^-}({\rm min}) = \cos^{-1} [ -1 + 2 \beta^2],
\label{min}
\end{eqnarray}
as was obtained earlier by Barducci and Toni in ref.~\cite{Bar23}.

The opening angle between $e^+$ and $e^-$ in  ATOMKI experiments 
has been measured
in the laboratory system.  
Since  the ATOMKI experiments deal with proton reactions with beam energies  
of only a few MeV,  which are much smaller than the proton mass of about 1 GeV,  the 
difference in the opening angles in the laboratory system and the CM system are small.
It suffices  to compare   the opening angle with experimental data measured in the laboratory system with 
theoretical results calculated in the CM system, using Equation~(48), without incurring serious errors.

We study here the minimum opening angles
$\theta_{e^+e^-}$(min) in terms of $\beta^2$ and experimental
kinematic attributes in different $AB$ fusion collisions so that we can study
the emission of the $X$ particle in the de-excitation process, not only
to the ground state $C_{\rm gs}$ but also
to an excited state $C_{\rm f}$ of the compound nucleus.   The quantity 
$\beta^2$ in Equation (\ref{min})  by
\begin{eqnarray}
\beta^2=\vec \beta^2 = \frac{2 K}{m_X} 
 \left [1+
 \frac{K}{2m_X}
 \right ] \biggr / 
 \left \{ 1 + \frac{2 K}{m_X} 
 \left [ 1+
 \frac{K}{2m_X}
 \right ]  \right \},
\end{eqnarray}
where  $K$ is  the kinetic energy of $X$ and its emission  partner $C_{\rm f}$ in the CM frame.
  The quantity $K$ is given explicitly by 
\begin{eqnarray}
&&K=  E_x  -[M(C_{\rm f})-M(C_{\rm gs})] -m_X,
\\
&&E_x=  \frac{AB}{A+B} E_A^{\rm lab}+  Q_{\rm gs} ,
\\
&&Q_{\rm gs}=  M(A) + M(B)-M(C_{\rm gs}),
\label{last}
\end{eqnarray}
where $E_x$ is the excitation energy of the compound nucleus $C^*$
relative to its ground state $C_{\rm gs}$; $M(C_{\rm f})$ and
$M(C_{\rm gs})$ are the masses of the nucleus $C$ at the $C_{\rm f}$  state and
the ground state $C_{\rm {gs}}$, respectively; $Q_{\rm gs}$ is the $Q$ value of the
$AB$ fusion reaction relative to the ground state $C_{\rm gs}$ of the
fused compound nucleus $C$; and $M(A)$ and  $M(B)$
are the masses of the projectile and target nucleus $A$ and $B$ with atomic mass numbers $A$ and $B$, respectively. 

 The quantity $K$ is the sum of
the kinetic energy of X and the kinetic energy of its emission partner
$C_{\rm f}$ in the CM system,
 \begin{eqnarray}
K = \sqrt{|\vec p_X|^2 + m_X^2} - m_X + \frac{|\vec p_X|^2} {2 M(C_{\rm f})}.
\end{eqnarray}

As $ M(C_{\rm f}) \gg m_x$ and $|\vec p_X|^2/2m_X \gg |\vec
p_X|^2/2M(c_{\rm f})$, the quantity $K$ is essentially the kinetic
energy of X in the CM system.

We specialize to ATOMKI experiments, where
  we have
\begin{eqnarray}
&&Q_{\rm gs}(p + {}^3{\rm H}  \to {}^3{\rm He}_{\rm gs})= M(p)+M({}^3{\rm H}) - M(^4{\rm He}_{\rm gs})= 19.815 ~{\rm MeV},
\\
&&Q_{\rm gs}(p + {}^7{\rm Li}  \to {}^8{\rm Be}_{\rm gs})= M(p)+M({}^7{\rm Li}) - M(^8{\rm Be}_{\rm gs})=  17.255 ~{\rm MeV},
\\
&&Q_{\rm gs}(p + {}^{11}{\rm B}  \to {}^{12}{\rm C}_{\rm gs})= M(p)+M({}^{11}{\rm B}) - M(^{12}{\rm C}_{\rm gs}) =  15.957~ {\rm MeV}.
\end{eqnarray}

The $X$ particle can be
presumed to be the hypothetical X17 particle or the isoscalar QED
meson.   While the internal pair conversion contributions to the
$\theta_{e^+e^-}$ distribution are smooth functions of
$\theta_{e^+e^-}$, the decay of $X17$ into $e^+e^-$ gives a relatively
sharp delimiter to allow a reasonable extraction of the experimental
$\theta_{e^+e^-}({\rm min})$ data.  From the ATOMKI data \linebreak  in~\cite{Kra16,Kra21,Kra22,Kra23}, we extract the minimum opening angle $\theta_{e^+e^-}({\rm min})$ for
different combinations of collision targets, energies, and
de-excitation possibilities as listed in~Table~\ref{tb2}.

\begin{table}[h]
\centering
\caption { Experimental $\theta_{e^+e^-}$(min) data
for different collision targets, energies, and final state possibilities  extracted from the ATOMKI data in  \cite{Kra16,Kra21,Kra22,Kra23}. 
Here $E_x$ is the excitation energy of the compound nucleus
  $C^*$    relative to
  the compound nucleus ground state $C_{\rm gs}$, after proton fusion 
  in the $p+A\to C^*$ reaction.   The quantity $K$ is the X17
  kinetic energy in the center-of-mass system.  We use here
  $m_X=16.70$ MeV \cite{Kra21} in the evaluation of $K$.}
\vspace*{0.3cm}
\begin{tabular}{|c|c|c|c|c|c|c|c|}
\cline{1-6}
 \multicolumn{1}{|c|}{\!\!Reaction}&  $E_p^{\rm lab}$ &\!\!Compound\!\!& $E_x$ & $K$   & $\!\theta_{e^-e^+}$(min) \!\!\!\!\\
 \multicolumn{1}{|c|}{} & (MeV) &nucleus  & (MeV) &  (MeV)  & (degree)\\ 
\hline
$p$+$^3$H  & 0.510 \cite{Kra21} &$^4$He$^*$ &  20.20   & 3.50 $\pm$ 0.5 & 100$\pm$5\\ 
                        & 0.610 \cite{Kra21} &$^4$He$^*$ & 20.27  & 3.57 $\pm$ 0.5 &   90$\pm$5 \\ 
                        & 0.900 \cite{Kra21} &$^4$He$^*$ & 20.49  & 3.79 $\pm$ 0.5&   96$\pm$5\\  
\hline
$p$+$^7$Li & 1.10 \cite{Kra16,Kra21}& $^8$Be$^*$ &  18.22 & 1.52 $ \pm $ 0.5 & 130 $\pm$ 5\\ 
\hline
         $p$+$^7$Li                &      4.0 \cite{Kra23}                           &    $^8$Be$^*$                 &     &  $^8$Be$^*$$\to$$^8$Be(gs)+X17 &          \\ 
 &   &  &  20.76 &         4.06 $\pm$ 0.5          &     110 $\pm$ 5                    \\ 
                        \hline

          $p$+$^7$Li                &       4.0 \cite{Kra23}                                &        $^8$Be$^*$             &      &\!\!\!$^8$Be$^*$\!\!$\to$$^8$Be$(2_1^+\!\!)$(3.03MeV)+X17\!\!\!\!& 
                          \\ 
                                                 &                                &                         &  20.76   &      1.03  $\pm$ 0.5          &    136 $\pm$ 6                      \\ 
\hline
 
$p$+$^{11}$B & 1.50 \cite{Kra22}& $^{12}$C$^*$ &  17.33 & 0.63 $ \pm $ 0.5 & 145 $\pm$ 3\\ 
                             & 1.70 \cite{Kra22}& $^{12}$C$^*$ &  17.52 & 0.82 $ \pm $ 0.5 & 144 $\pm$ 3\\ 
                            & 1.88 \cite{Kra22}& $^{12}$C$^*$ &  17.68 & 0.98 $ \pm $ 0.5 & 138 $\pm$ 3\\ 
                            & 2.10 \cite{Kra22}& $^{12}$C$^*$ &  17.88 & 1.18 $ \pm $ 0.5 & 134 $\pm$ 3\\ 
\hline
\end{tabular}
\label{tb2}
\end{table}

\begin{figure}[h]
\centering
\includegraphics[scale=0.40]{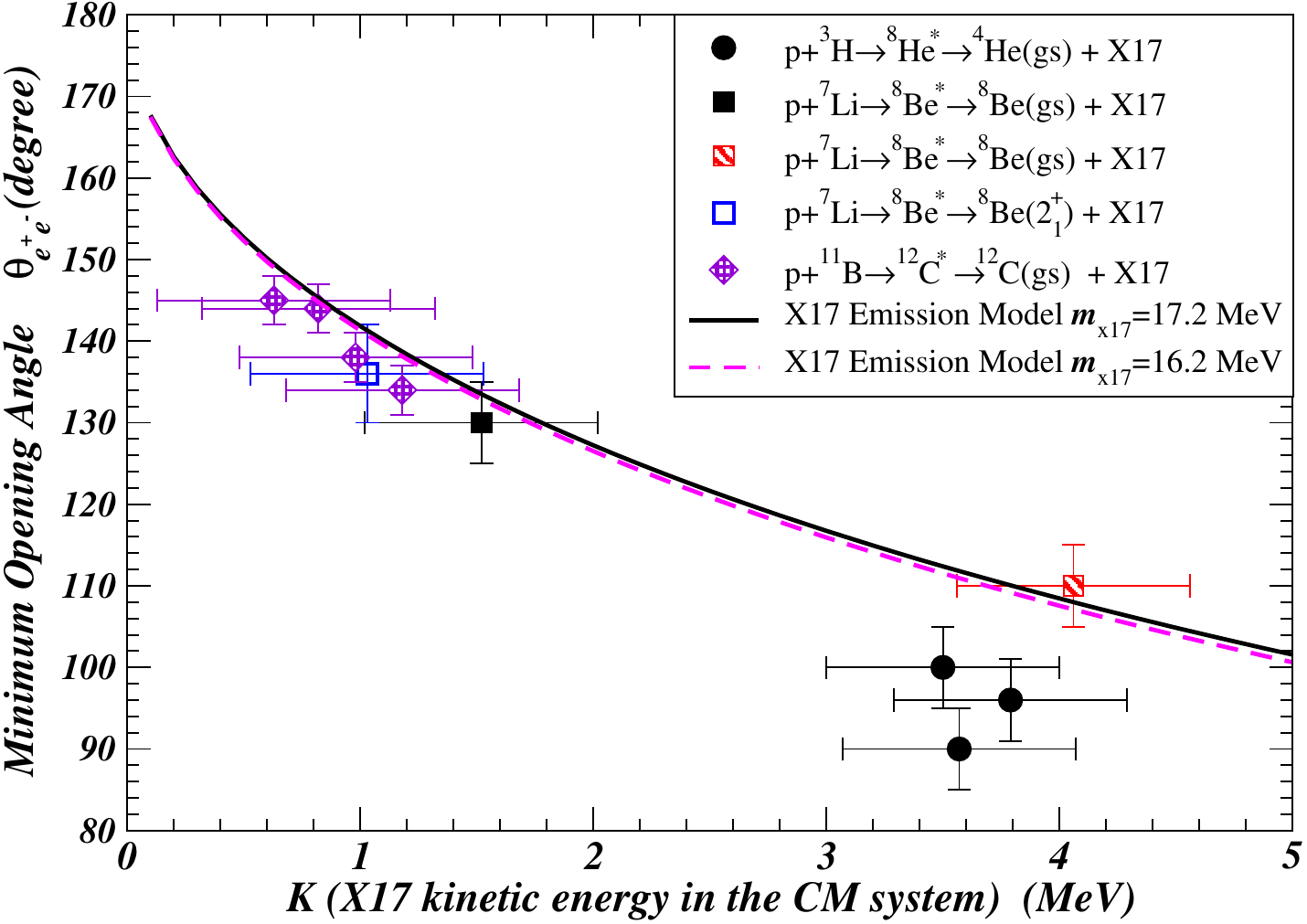}
\caption{{Comparison of the ATOMKI} 
 experimental data~\cite{Kra16,Kra21,Kra22,Kra23} with  the ATOMKI X17 emission model 
predictions of the minimum opening angle  $\theta_{e^+e^-}$(min) between $e^+$ and $e^-$
 as a function of the X17
kinetic energy $K$ in the CM frame, for different collision energies, targets, and final states.
The X17 emission model envisages the fusion
  of the incident proton $p$ with the target nucleus $A$, forming a
  compound nucleus $C^*$, which subsequently de-excites to the
  final state $C_{\rm f}$ with the
  simultaneous emission of the X17 particle.  The subsequent decay of the
  X17 particle into $e^+$ and $e^-$ then gives the angle
  $\theta_{e^+e^-}$ between $e^+$ and $e^-$.  The solid curve is the theoretical prediction
  of  $\theta_{e^+e^-}$(min)
 as a function of the X17 kinetic energy $K$, and the ATOMKI data  points are from ~\cite{Kra16,Kra21,Kra22,Kra23} as
 summarized in~Table~\ref{tb2}.}
 \label{fig4}
\end{figure}

  In the extraction of $\theta_{e^+e-}$\!(min) from the
experimental ATOMKI data, 
although there may be some ambiguities and
uncertainties,
we choose the $\theta_{e^+e^-}$(min) value to be
the midpoint between a sudden jump in the angular distribution of
$\theta_{e^+e^-}$ at the onset of the anomaly.  
The uncertainty of the value of the kinetic energy $K$ of the X particle in Table~\ref{tb2} comes from the uncertainty of the X17 mass, which  is of the order 0.5 MeV, and the uncertainty in the $\theta_{e^+e^-}$(min) is taken to be the grid size of the measurement.  All final nucleus
   states $C_{\rm f}$ in Table~\ref{tb2} are implicitly  ground states, except when specified explicitly as for the case of $p+^7$Li at $E_p^{\rm lab}$ = 4 MeV.
The experimental
attributes of projectile nucleus, target nucleus, proton collision
energies, and final compound nucleus states $C_{\rm f}$ furnish sufficient
information to allow the determination of the important physical
quantity $K$, the X17 kinetic energy in the CM frame.  To compare the
theoretical predictions of the X17 emission model with experimental
data, we plot in Figure~\ref{fig4} the minimum opening angle
  $\theta_{e^+e^-}$(min) as a function of $K$.  The
  theoretical predictions  from
  Equations (\ref{min})--(\ref{last}) are given as the solid curve.  
  One
  observes that there is  a reasonable agreement between the theoretical
curve and the data in Figure~\ref{fig4}, indicating the validity of the X17 emission model as was first suggested
by the ATOMKI Collaboration~\cite{Kra16}.  The theoretical curve is insensitive to the change of the X17 mass, and the
agreement persists when the X17 mass changes from 16.2 MeV to 17.2 MeV as shown in Figure~\ref{fig4}.

The minimum opening angle can be used to delimit the region of
the $e^+e^-$phase space relevant for the production of the X17.  In the
evaluation of the invariant mass of the $e^+e^-$ pair, the
$\theta_{e^+e^-}$(min) delimiter can be used to exclude contributions
of $e^+e^-$ yields with opening angles much less than
$\theta_{e^+e^-}$(min), which can help eliminate part of the $e^+e^-$
background.  It indicates that the X17 kinetic energy variable $K$
in the CM frame is the most important kinematical variable in setting
the limit of the opening angle.  The correlation of the experimental
$\theta_{e^+e^-}$(min) with other reaction attributes will provide
useful information on the associated likelihood of X17 production.
For example, one may ask how $\theta_{e^+e^-}$(min) correlates with
$K$, when $K$ is allowed to vary between resonances.  Is the resonance
with particular quantum numbers essential for the presence or absence
of $\theta_{e^+e^-}$(min)?

ATOMKI reported the observation of  the emission of the   X17 particle
 in 
 the de-excitation of the compound
nucleus ${}^8$Be$^*$ to the excited state of ${}^8$Be(3.03 MeV)~\cite{Kra23}.   We show here in Fig.\ 4 that the minimum opening angle 
$\theta_{e^+e^-}$(min) in  the de-excitation to the excited state
follows the same systematics as other reactions with the X17 emission.  The  ATOMKI X17 emission model
is therefore shown to be valid also for the de-excitation of the compound nucleus $C^*$ to an excited state $C_{\rm f}$ of the  compound nucleus.
Such a case occurs at a higher collision energy  ($p_p^{\rm lab}=4$ MeV), leading to the
population of a higher compound nucleus state.  Such a method of X17
production from a higher excited state $C^*$ to a lower excited state
$C_{\rm f}$ of the compound nucleus opens the door for X17 production
at higher fusion collision energies and in fusion reactions using different (projectile A)-(target B) 
combinations.  Furthermore, there is a relatively large X17 yield in the 
backward angle region near $\theta_{e^+e^-}$=$\pi$, where there may be 
a relatively low competing
$e^+e^-$ background.   It
may provide a way to study the properties
of the produced parent X17 particle.
  It opens up a valuable avenue for future
investigation of the X17 particle.

The magnitudes of $K$ in Table~\ref{tb2} provide useful information on the
angular momentum carried by the X17 particle when it is emitted
simultaneously with its emission partner $C_{\rm f}$ by the compound
nucleus $C^*$ in the CM frame.  The quantity $K$ in Table \ref{tb2} covers a
range of only a few MeV.  The small magnitude of $K$ suggests that
if $X$ is a point particle, then
 in
the emission of $X$ by the compound nucleus $C^*$, the emission  will occur dominantly
with $X$ in the $S$-wave in the CM frame.  The emission of
X17 carrying $l=1$ will need to overcome a centrifugal barrier of
order $E_l\sim\hbar^2 l(l+1)/2 \mu R^2$ where $\mu=m_X m_{C^*}/(m_X +
m_{C^*}) \sim m_X$.  The barrier $E_l$ is $\sim$23 MeV even for $R\sim
10$ fm, much greater than the $K$ values in Table~\ref{tb2}.  On the other hand, if the $X$ particle is a spatially  extended object or 
a wave phenomenon,
then the angular momentum may  be carried at  larger separations outside the nucleus.    The 
spatial nature of the $X$ particle and the  angular momentum distribution of the X particle at the moment of emission  will need to be investigated.    Models of the $X$ emission with non-zero $l$ values will need to be consistent 
with the small value of 
the kinetic energy $K$.

\subsection{The Dubna Observation of the X17 and the E38 Particle by Diphoton Measurements}

\begin{figure} [h]
\includegraphics[scale=0.50]{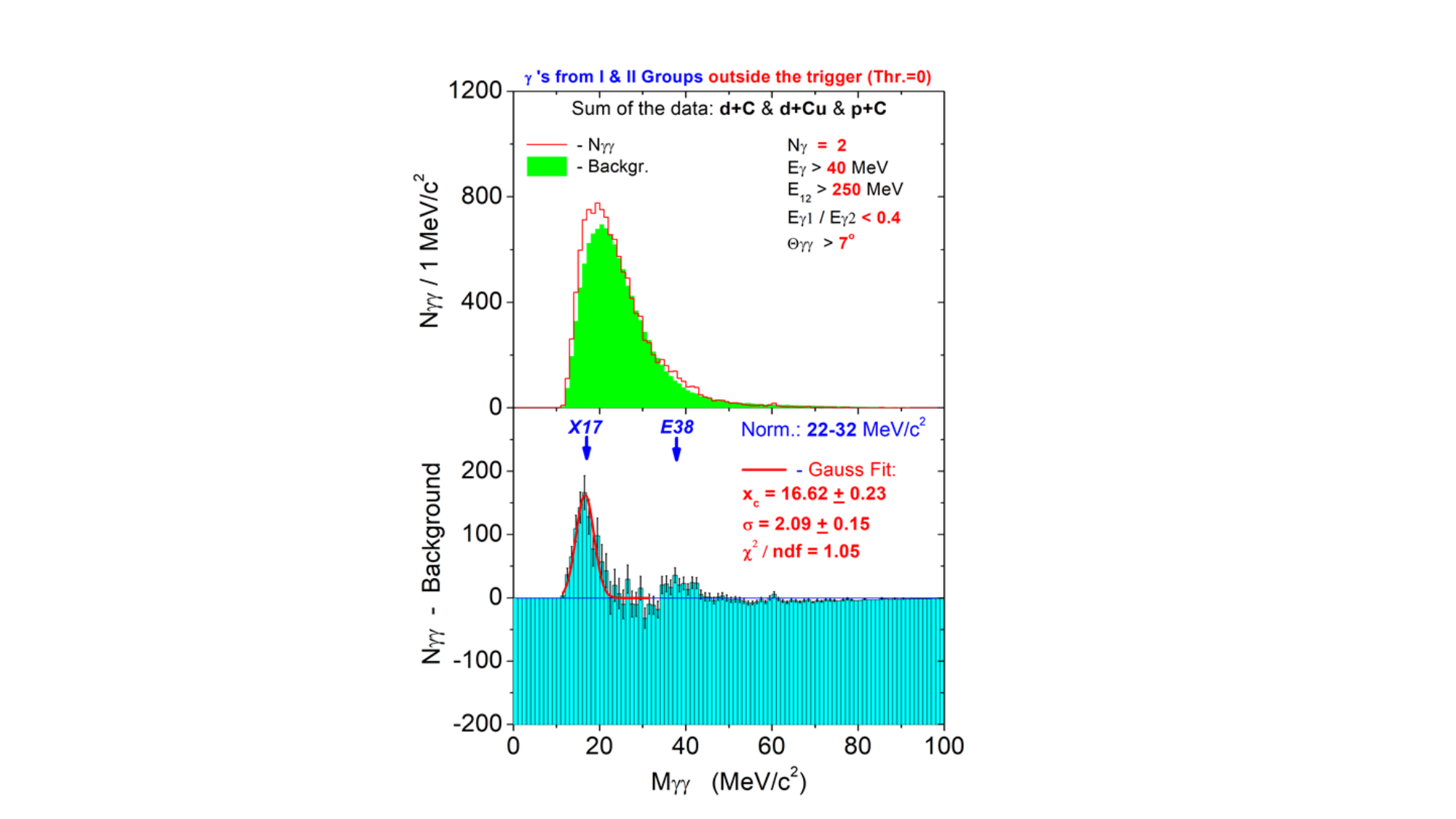}
\caption{{The diphoton} 
 invariant mass spectra from light ion
  collisions with C and Cu targets at a few GeV per nucleon at the
  JINR Nuclotron, Dubna~\cite{Abr23}.  The solid curve in the upper
  panel shows the invariant mass distribution obtained by combining
  two photons from the same event, and green shaded region the
  invariant mass distribution by combining two photons from mixed
  events.  The signal of correlated photons subtracting the mixed
  event background gives the signal  represented by the blue region in the lower panel, where the
  resonance-like structures at $\sim$17 and $\sim$38 MeV show up.}
  \label{fig5}
  \end{figure}

Abraamyan and collaborators at Dubna have been investigating the
two-photon decay of particles to study the resonance structure of the
lightest hadrons near their thresholds, using $d$ and $p$ beams of a
few GeV with fixed internal C and Cu targets at the JINR Nuclotron~\cite{Abr09,Abr12,Abr19,Abr23}.  Their PHONTON2 detector consists of
two arms placed at 26 and 28 degrees from the beam direction, with
each arm equipped with 32 lead-glass photon detectors.  The photon
detectors measure the energies and the emission angles of the photons,
from which the invariant masses of the photon pairs can be measured.
By selecting photon pairs from the same arm with small opening angles,
it is possible to study neutral bosons with small invariant masses,
such as those below the pion mass gap $m_\pi$.  They reported earlier
the observation of a resonance structure at a mass of $\sim$38~MeV~\cite{Abr12,Abr19}.  In a recent analysis in the diphoton spectrum
extended down to the lower invariant mass region, the Dubna
Collaboration reported the observation of resonance-like structures
both at $\sim$17 and $\sim$ 38 MeV in the same experimental setup, in
support of the earlier ATOMKI observation of the hypothetical X17 particle
and the earlier Dubna observation of the hypothetical E38 particle~\cite{Abr23}.  The resonance structure of the diphoton signal of the X17
particle appears to be quite strong and prominent as shown in Figure~\ref{fig5}.

The observation of X17 and E38 at Dubna completes an important piece
of the anomalous particle puzzle, as the isoscalar X17 and the
isovector E38 come in a pair, and they are orthogonal linear
combinations of the $|u\bar u\rangle $ and $|d \bar d\rangle$
components.  The agreement of their masses with those predicted by the
phenomenological open-string model of QED-confined $q\bar q$ model~\cite{Won10,Won20} lends support to the description that
a quark and an antiquark can be confined and bound as stable
QED-confined mesons interacting in the Abelian U(1) QED interaction.
This is a rather unusual and unfamiliar concept.  The confirmation of
these anomalous particle observations will be therefore of great
interest.

\begin{figure} [h]
\centering
\vspace{-6pt}
\includegraphics[scale=0.45]{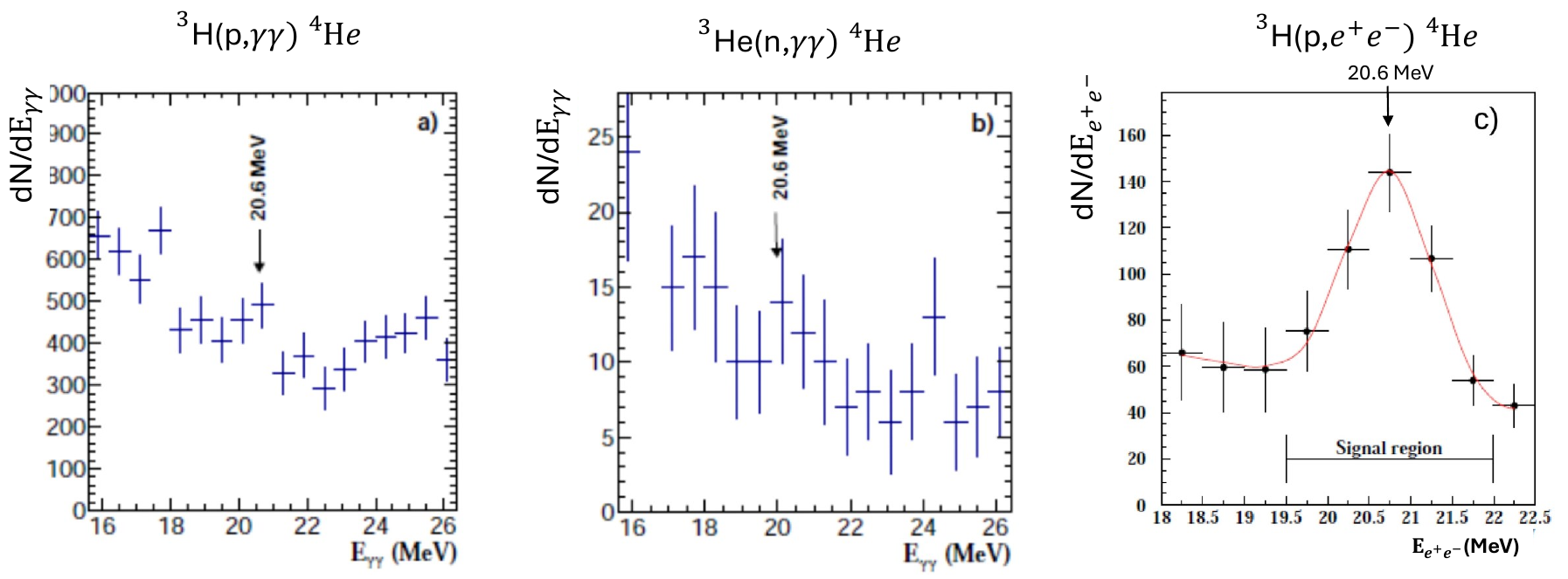}
\caption{(\textbf{a}) {The spectrum of the diphoton} 
 energy sum, $dN/dE_{\gamma \gamma}$, in
  the  $^3$H$(p,\gamma\gamma){}^4$He$_{\rm gs}$ reaction with a proton
  beam energy at 1 MeV~\cite{Nag19}, (\textbf{b}) the spectrum of the diphoton  energy sum, $dN/dE_{\gamma
    \gamma}$, in the $^3$He$(n,\gamma\gamma){}^4$He$_{\rm gs}$
  reactions using a cold neutron beam line~\cite{Nag19}, and  (\textbf{c}) the 
  spectrum of the dilepton energy sum,  $dN/dE_{e^+e^-}$,  in the $^3$H$(p,e^+e^-){}^4$He$_{\rm
    gs}$ reaction with a proton beam energy at 0.9 MeV, where the decay of X17 to $e^+$ and $e^-$ has been observed~\cite{Kra21}. }
  \label{fig10}
\end{figure}

The Dubna observation of the diphoton invariant mass at $\sim$17 MeV
supports the ATOMKI experimental finding of the hypothetical X17
particle using the $e^+e^-$ decay.  It suggests further that such a
diphoton decay should also occur for the X17 particle produced in
ATOMKI-type experiments.  
In the ATOMKI--Krakow--Munich
Collaboration experiments, 
Nagy et~al. 
search for the  diphoton
decays of the X17 particle.  In the $^3$H($p,e^+e^-)^4$He$_{\rm
  gs}$ experiment using the {2MV tandem Accelerator of MTA}, a proton
beam at 1 MeV collides onto a $^3$H target, and the gamma rays are
detected in coincidence with 14 LaBr scintillates.  The spectrum of 
the diphoton energy sums, 
 $dN/dE_{\gamma \gamma}$,  from the $^3$H$(p,\gamma\gamma)^4$He$_{gs}$ reaction, is  shown in Figure~6a.  In another
experiment with the $^3$He($n,\gamma\gamma)^4$He$_{\rm gs}$ reaction
using the cold neutron beam line of the high-flux reactor at Research
Neutron Source of the Technical
University of Munich, the $Q$ value of the reaction is 20.6 MeV, so the
resonance absorption populates the $E_x = 20.21$ MeV state as well as the
$E_x$ = 21.01 MeV $^4$He$^*$ state using a ${}^3$H gas target.  The
photons are detected in coincidence by an array of 12 scintillators.
The spectrum of the diphoton energy sum, $dN/d E_{\gamma \gamma}$, 
from the $^3$He$(n,\gamma\gamma)^4$He$_{gs}$ reaction
is shown
in Figure~\ref{fig10}b.  As stated in~\cite{Nag19} for these measurements, ``a
peak clearly shows up at 20.6 MeV'' in the $^3$H($p,\gamma
\gamma){}^4$He reaction in Figure~6a,
but
the statistics is, however, poorer.

It is interesting to compare the spectrum of the diphoton energy sum,  $dN/dE_{\gamma \gamma}$,
with the spectrum of the
dilepton energy sum, 
$dN/dE_{e^+e^-}$,  in Figure~\ref{fig10}.   The $e^+$ and $e^-$ particles from the X7 decay
have highly relativistic energies, and thus the $e^+e^-$ emission from the X17 dilepton decay 
follows similar  kinematics as the $\gamma \gamma$ emission from  the  X17 diphoton decay,  and vice
versa.    If X17  decays into $e^+e^-$  and  also into $\gamma\gamma$, then
 $dN/dE_{e^+e^-}$ and  $dN/dE_{\gamma\gamma}$  should be very similar
 for reactions with similar experimental conditions.
In Figure~\ref{fig10}c,  we show $dN/dE_{e^+e^-}$ in 
 the $^3$H$(p,e^+e^-){}^4$He$_{\rm
    gs}$ reaction with a proton beam of similar energy at 0.9 MeV, where the decay of X17 to $e^+$ and $e^-$ has been observed.
The $dN/dE_{e^+e^-}$ 
spectrum in {Figure~\ref{fig10}(c)} 
exhibits a peak at $E_{e^+e^-}$$\sim$20.6 MeV, for which the most likely opening angle 
$\theta_{e^+e^-}$  is 120$^{\rm o}$~\cite{Kra21} and  corresponds  to an invariant mass $m_{e^+e^-}$ of the $e^+e^-$ pair to be  
\vspace{-6pt}
\begin{eqnarray}
(m_{e^+e^-})^2 =  2 E_{e^+}  E_{e^-} (1-\cos \theta_{e^+e^-})  
\sim  2 \left (\frac{E_{e^+} + E_{e^-}}{2} \right )^2(1-\cos120^{\circ}) = (17.8 {\rm MeV})^2,
\end{eqnarray}
pointing to the observation of the X17 particle at about 17.8 MeV. 
The coincidence of 
 the peak position  of the  spectra for the dilepton energy sum $dN/dE_{e^+e^-}$  in Figure~\ref{fig10}c with 
the peak positions  of the  spectra for the diphoton energy sum $dN/dE_{\gamma \gamma}$ in Figure~\ref{fig10}a,b
 indicates the likely occurrence of X17 diphoton decay in ATOMKI-type experiments,  consistent with the X17 diphoton decay observed by  the Dubna Collaboration.  
The statistics of the peaks in the diphoton energy sum  spectra  in Figure~\ref{fig10}a,b in ~\cite{Nag19} is, however, poor, and it is important to confirm these  diphoton measurements with better statistics to demonstrate the occurrence of X17 diphoton decay in ATOMKI-type experiments.

It is worth noting that 
the mathematical formulation 
in the X17 emission model to
calculate  the opening angle  between $e^+$ and $e^-$  is the same as the 
formulation to calculate  the opening angle between the two photons in the diphoton  decay of X17.
Consequently, the distribution of the opening angle between the two photons  for the X17 diphoton decay
 would have the same shape as 
 the  distribution of the opening angles between $e^+$ and $e^-$ in the X17 dilepton decay.   The dependence of the minimum opening angle between the two photons in the CM system on the X17 kinetic energy $K$ 
 should likewise be the same as that between $e^+$ and $e^-$.   The extensive set of ATOMKI data on X17 
 dilepton decay into 
 $e^+$ and $e^-$ in various reactions
 may be a useful resource in predicting the behavior for the X17 diphoton decay, which may  assist in the design of diphoton measurements 
 in proton \linebreak  fusion~experiments.
 
Possible direct X17 and E38 diphoton decay signals may come
from the resonance structure of the diphoton invariant mass at around
15 MeV and 38 MeV in the $pp \to pp \pi^+ \pi^-(\gamma \gamma) $
reaction at $p_{\rm lab}(p)$= 190 GeV/c~\cite{Ber11,Ber14,Sch11,Sch12,Ber12}
and the $\pi^- p \to \pi^- p_{\rm slow} \pi^+ \pi^-(\gamma \gamma) $
reaction at $p_{\rm lab}(\pi^-)$ = 190 GeV/c, obtained by the COMPASS
Collaboration as pointed out by~\cite{Bev11,Bev12,Bev12a,Bev20}.  A
possible direct signal may also be the prominent resonance structure
at 38 MeV in the diphoton invariant mass spectrum in PbPb collisions
at $\sqrt{s_{_{\rm NN}}}=2.76$ TeV (Figure~5.6 in ref.~\cite{Sno12})
obtained by the CMS Collaboration~\cite{CMS13}.

\section{Implications of Quark Confinement in the QED Interaction}

\subsection{Confinement May Be an Intrinsic Property of Quarks} 

The observations of the anomalous soft photons, the X17 particle, and
the E38 particle provide promising evidence for the possible existence
of the QED mesons, which are hypothetical $q\bar q$ states confined
and bound  non-perturbatively  by the QED interaction.  Their masses at
$\sim$17 and $\sim$38 MeV are close to the theoretically predicted
masses of isoscalar and isovector QED mesons.  The occurrence of the
isoscalar and isovector doublet reflects properly the two-flavor
nature of the light quarks.  Their decays into $\gamma \gamma$ and
$e^+e^-$ indicate their composite nature and their connection to the
QED interaction.  Their modes of production by low-energy
proton fusion and by high-energy nuclear collisions can also be
understood in terms of the production of quark--antiquark pairs by soft
gluon fusion or the $(q\bar q)$ production by string fragmentation in
high-energy hadron--hadron collisions~\cite{Won20}.

While the confirmation of the observations is pending, it is of
interest to examine the implication of the existence of the QED mesons
if these observations are indeed confirmed under further  scrutiny.  It will
indicate that the attribute of quark confinement may not be the sole
property of the QCD interaction alone and that quarks may also be
confined in the QED interaction as QED mesons in the mass region of
many tens of MeV.  It is possible that the confinement attribute may
be an intrinsic property of the quarks.  
Such a possibility is consistent
with the observational absence of free quarks or fractional charges.
It indicates further that in the multitudes of non-perturbative
interactions between a quark and an antiquark, there may be an
underlying quark confinement principle which holds that in the
dynamics of quarks in different interactions, each interaction always
leads to the confinement of quarks.  A quark and its antiquark may be
confined and bound as a neutral boson in the weak interaction and the
gravitational interaction with the exchange of a $Z^0$ boson or a
graviton.
 
The possibility of quarks being confined in the QED interaction also
implies that the QED interaction between a quark and an antiquark
may differ from the QED interaction between an electron and a positron.
It will be of great interest to inquire what other additional
different properties there can be between quarks and electrons and why
they are different.  For example, the integer electric charge of electrons and fractional electric
charges of quarks remain an unresolved problem.  There
is the question of whether the QED interaction between an electron and a
positron may belong to the non-compact non-confining QED theory while
the QED interaction between a quark and an antiquark belong to the
confining compact QED theory.  The possibilities of the compact and
non-compact QED bring with them the question whether the QED
interaction is unique or is endowed with a multitude of
experimentally testable particle-dependent possibilities with
different properties.  A related question is whether the QED
interaction between quarks in a nucleon may also contain the linear
QED confinement component that depends on the magnitudes and signs of
the electric charges in addition to the standard Coulomb component.

\subsection{QED Meson Assembly and Dark Matter} 

Astrophysical objects consisting of a large assembly of isoscalar
$I(J^\pi)=0(0^-)$ QED mesons such as the X17 particle with a mass
$m_X$ = 17 MeV will be  electron--positron emitters, gamma ray emitters, or dark
black holes with no emission.  The mode of emission, the emission
energies, and the lifetimes depend on the gravitational energy of the
assembly.  Such assemblies of QED mesons present themselves as good
candidates of $e^+$$e^-$ emitters, gamma-ray emitters, or the
primordial dark matter.  We can make estimates on the constraints on
masses and radii of such assemblies where they may be found.

If we consider an assembly of $A$ number of $m_X$ QED mesons of mass
$M_A$ $\equiv$ $M$ and we place a test QED $m_X$ meson at the surface of
the assembly at radius $R$, the mass $M_{A+1}$ of the combined system
is
\vspace{-6pt}\begin{eqnarray}
M_{A+1} = M_A + m_X - \frac{G M_A m_X}{Rc^2},
\end{eqnarray}
where $G$ is the gravitational constant.  The $Q$ value for the test
QED meson at the surface of the ($A$+1) assembly to decay into an
electron--positron pair is
\begin{eqnarray}
\hspace*{-0.7cm} Q( (A+1) \!\to\! A+e^+e^-)=\! m_X c^2\! -\!\frac{G
  M_{A} m_X}{R}\! -\! 2m_e c^2.
\end{eqnarray}

The $Q$ value for the test QED meson to decay into two photons is
\begin{eqnarray}
Q((A+1) \to A +2\gamma)= m_X c^2 -\frac{G M_{A} m_X}{R}.
\end{eqnarray}

{Thus,  the QED meson assembly of mass $M$ and radius $R$ will be an $e^+e^-$ and $\gamma$ emitter~if  }
\begin{eqnarray}
\frac{M}{ R} < \frac{c^2}{G}\left (1-\frac{2m_e}{m_X} \right ).
\end{eqnarray}

The QED meson assembly will emit only gamma rays but no $e^+$$e^-$
pairs if
\begin{eqnarray}
 \frac{M }{R } > \frac{c^2}{G} \left (1-\frac{2m_e }{m_X}
\right ).
\end{eqnarray}

The QED meson assembly will not emit $e^+$$e^-$ pairs nor gamma rays, and is a dark assembly of matter 
if
\vspace{-6pt}
\begin{eqnarray}
  \frac{M }{R } > \frac{c^2}{G},
\label{MM4}
\end{eqnarray}
which is essentially the condition for a QED meson black hole.

As the evolution of the earlier Universe is likely to have passed
through the phase of the quark--gluon plasma and quarks are essential
constituents of the quark--gluon plasma, low-lying confined
quark--antiquark states such as the isoscalar QED meson may play
important roles in the states of matter after the phase transition
from the quark--gluon deconfined phase to the confined matter phase.
Astrophysical objects consisting of a large assembly of QED meson
$q\bar q$ states may therefore be of interest.  It offers a pathway
for electron--positron emitter, gamma emitter, and black hole formation
through the quark--gluon plasma deconfinement phase that is different
from the evolutionary path of black hole formation by way of the
baryonic stellar evolution.

\subsection{New Family of QED-Confined Particles and Dark QED Neutron} 

The QED mesons are composite objects with a complex structure.  They
possess additional degrees of freedom of spin--spin, spin--orbit,
collective vibrational and rotational motion, flavor admixture, and
molecular excited QED mesons states.  For example, we can get some
idea of the vibrational states from the spectrum of a stretched string
as shown in Figure~7 of~\cite{Won22}.  The possibility of adding quarks
with different flavors, angular momentum, and spin quantum numbers
will add other dimensions to the number of species of the QED-confined
$q\bar q$ composite particles.

The success of the open-string description of the QCD and QED mesons
leads to the search for other neutral quark systems stabilized by the
QED interaction between the constituents in the color-singlet
subgroup, with the color-octet QCD gauge interaction as a spectator
field.  Of particular interest is the QED neutron with the $d$, $u$,
and $d$ quarks~\cite{Won22,Won22a}.  They form a color product {group
of} ${\bb 3}$ $\otimes$ $ {\bb 3} $ $\otimes$ $ {\bb 3}$ = ${\bb 1}
\oplus {\bb 8} \oplus {\bb 8} \oplus {\bb {10}}$, which contains a
color-singlet subgroup $\bb 1$, where the color-singlet currents and
the color-singlet QED gauge fields reside.  In the color-singlet
$d$-$u$-$d$ system with three quarks of different colors, the attractive QED
interaction between the $u$ quark and the two $d$ quarks overwhelms
the repulsion between the two $d$ quarks to stabilize the QED neutron
at an estimated mass of 44.5 MeV~\cite{Won22}.  The analogous QED
proton has been found to be theoretically unstable because of the
stronger repulsion between the two $u$ quarks, and it does not provide
a bound state nor a continuum state for the QED neutron to decay onto
by way of the weak interaction.  Hence, the QED neutron may be stable
against the weak interaction.  It may have a very long lifetime and
may be a good candidate for the dark matter.  Because QED mesons and
QED neutrons may arise from the coalescence of deconfined quarks
during the deconfinement-to-confinement phrase transition in different
environments, such as in high-energy heavy-ion collisions, neutron star
mergers~\cite{Bau19,Bau20,Wei20}, and neutron star cores~\cite{Ann20},
the search for the QED bound states in various environments will be of
great interest.

\subsection{Beyond the Confining Interaction of a Quark and an Antiquark in (3+1)D}

A quark and an antiquark reside predominately in (1+1)D, in which the
interaction between a quark and an antiquark is the linear confining
interaction for both QED and quasi-Abelian QCD as discussed in
Sections~\ref{sec3}~and~\ref{sec4}.  In the physical (3+1)D space-time, such a linear
interaction is only the dominant part of the full interaction between
the quark and the antiquark.  There will be additional residual
interactions arising from the presence of the transverse degrees of
freedom.  There are also contributions from the spin--spin, spin--orbit,
tensor, and other higher-order terms of the Breit interaction~\cite{Bar92,Won00,Won01,Cra09}.
 
For a confining string with a string tension $\sigma$, L\" uscher,~\cite{Lus81,Lus80} considered the fluctuations in the transverse
direction of the flux tube as a massless bosonic field theory in two
dimensions with a classical action, for which the action can be
integrated out to lead to a potential between a static quark at {${\bf
  r}_1$} 
 and an antiquark at {$\bf r_2$} in the large string length limit~as
\vspace{-6pt}\begin{eqnarray}
V( {\bf r}_1 {\bf r}_2)=\sigma |{\bf r}_1 - {\bf r}_2| +c  -  \frac{\alpha}{|{\bf r}_1 - {\bf r}_2|} + O\left (\frac{1}{{|\bf r}_1 - {\bf r}_2|^2}\right ),
\label{eq83}
\end{eqnarray}
where $\alpha$ depends on the coupling constant, and $c$ is a
constant.  These are therefore long-range residual interactions in
both the confined QCD and QED mesons.  They represent corrections that
arise from expanding the potential between a quark and an antiquark in
powers of their separation $ |{\bf r}_1 - {\bf r}_2| $.  A powerful
tool to study the non-perturbative behavior of the interquark
potential is the ``Effective String Theory'', in which the confining
tube contains the quark and the antiquark at the two ends
\cite{Nam70,Nam74,Got71,Lus81,Lus80,Pol91,Hel14,Aha13,Bon21}.  The
Nambu--Goto action can be integrated exactly in all geometries that are
relevant for lattice gauge theories: the rectangle (Wilson loop) in
ref.~\cite{Bil13}, the cylinder (Polyakov loop correlators) in
ref.~\cite{Lus04,Lus05}, and the torus (dual interfaces) in ref.~\cite{Bil06}.

For quarks with color charge numbers $Q_1^\qcd$ and $Q_2^\qcd$
interacting in the QCD interaction, we can match the above Equation~(\ref{eq83}) with the Cornell potential~\cite{Eic75} and the
phenomenological quark--antiquark potentials in ref.~\cite{Bar92,Won00,Won01,Won20,Cra09}.  Upon neglecting the
spin--spin, spin--orbit, other higher-order terms, and an unimportant
potential constant, we have the linear-plus-color-Coulomb interaction
of QCD
\begin{eqnarray}
V^\qcd( {\bf r}_1 {\bf r}_2)=  Q_1^\qcd Q_2^\qcd  \left [ \sigma^\qcd |{\bf r}_1-  {\bf r}_2|  - \frac{\alpha_s}{|{\bf r}_1 - {\bf r}_2|}    \right ] .
\end{eqnarray}

The quark and the antiquark also interact in the QED interaction.  We
can generalize the above to include both QCD ($\lambda=1$) and QED
($\lambda=0$) interactions to give
\begin{eqnarray}
V( {\bf r}_1 {\bf r}_2)&&=\sum_{\lambda=0}^1 (-1)^{\lambda+1} Q_1^\lambda Q_2^\lambda  \left [ \sigma^\lambda |{\bf r}_1-  {\bf r}_2|  -  \frac{\alpha_\lambda}{|{\bf r}_1 - {\bf r}_2|} 
\right ]t^\lambda, 
~ ~~ \lambda=\begin{cases} 0 & \text{QED} \cr 1  & \text{QCD} \cr \end{cases},
\label{eq85}
\end{eqnarray}
where $t^0$ is the generator of the U(1) gauge subgroup as defined in
Equation~(\ref{eq9}), and $t^1$ is a fixed generator of the SU(3)
subgroup in the eight-dimensional color-octet generator space in the
quasi-Abelian approximation of the non-Abelian QCD as discussed in
Section~\ref{sec3}.  The generators $t^0$ and $t^1$ satisfy $2{\rm
  tr}(t^\lambda t^{\lambda'}) = \delta^{\lambda \lambda'}$.

The above equation is for a single flavor.  In the case with many
flavors and flavor mixing, their effects can be taken into account by
replacing $Q_i^\lambda$ by the effective charge $\tilde Q_{i,{\rm
    eff}}^\lambda$ of Equation (\ref{27}).  It can be further generalized
to the case when the quark constituent and the antiquark constituent
possess different flavors.  For a composite $q_1 \bar q_2$ particle
with many flavors and flavor mixing, the above interaction between the
quark $q_q$ and the antiquark $\bar q_2$~becomes
\vspace{-6pt}\begin{eqnarray}
V( {\bf r}_1 {\bf r}_2)&&=\sum_{\lambda=0}^1 (-1)^{\lambda+1} \tilde Q_{q_1}^\lambda \tilde Q_{\bar q_2}^\lambda  \left [ \sigma^\lambda |{\bf r}_1-  {\bf r}_2|  -  \frac{\alpha_\lambda}{|{\bf r}_1 - {\bf r}_2|} 
\right ]t^\lambda.
\label{eq86}
\end{eqnarray}

When there is no flavor mixing as in the case of the charm and the
beauty quarks, the effective charge are just those of the standard
quark model, with $Q_{\{u,d,c,s,t,b\}}^\qcd$=1 and
$Q_{\{u,c,t\}}^\qed$=2/3, $Q_{\{d,s,b\}}^\qed$=$-1/3$, and $Q_{\bar
  q}^\lambda$=$ - Q_q^\lambda$.

\section{Conclusions and Discussions}

It has been observed that anomalous soft photons with transverse
momenta of many tens of MeV/c are proportionally produced when hadrons
are produced, and are not produced when hadrons are not produced,
indicating that the production of hadrons is always accompanied by
the production of neutral boson particles with masses in the region of
many tens of MeV.  Independently, in the search of axions with a mass
of many tens of MeV, an anomaly pointing to the production of a
hypothetical X17 neutral boson particle with a mass of about 17 MeV
has been observed in the decay of $^4$He, $^8$Be, and $^{12}$C excited
states at ATOMKI and also at the Hanoi University of Science.  There
have been also reported observations of the hypothetical X17 and the
hypothetical E38 particle at Dubna and at $\sim $17 and $\sim$38 MeV.  The
occurrence of these anomalies has led to the question  of whether quarks
may be confined when they interact non-perturbatively in the QED
interaction.

A related  question is whether there
are experimental circumstances in which a quark and an antiquark may
be produced and may interact non-perturbatively in QED interaction
alone, with the QCD interaction as an unexcited background
interaction.  We find that in hadron--hadron, $AA$, $e^+e^-$, and
$e^-A$ collisions, there can be situations in which a quark and an
antiquark may be produced with a center-of-mass energy below the pion
mass gap $m_\pi$ for collective QCD excitation, and the quark and its
antiquark can interact non-perturbatively in the QED interaction
alone, lest their non-perturbative interaction in the QCD interaction
endow the pair with a mass greater than or equal to the pion
mass.  It is therefore worth studying the question of quark
confinement in the QED interaction.

On the theoretical side, it is well known that according to the
Schwinger confinement mechanism, massless fermions interacting in the
Abelian QED gauge interactions in (1+1)D are confined for all
strengths of the gauge interaction, as in an open string, leading to a
confined and bound neutral boson with a mass proportional to the
magnitude of the coupling constant.

We ask next whether we can apply the Schwinger confinement mechanism
to light quarks.  Light quarks have masses of only a few MeV, and they
can be approximated as massless. A quark and an antiquark cannot be
isolated, so they reside predominantly in (1+1)D.  They can be
produced and interact in the QED interaction alone as we discussed
above.  The conditions under which the Schwinger confinement mechanism
can be applied are met when a light quark and a light antiquark are
produced and interact non-perturbatively with the QED interaction alone.  We can apply
the Schwinger mechanism to quarks to infer that a light quark and its
antiquark are confined in the QED interaction in (1+1)D.

On questions of quark confinement and QCD bound states, the
non-Abelian QCD interaction can also be approximated as a
quasi-Abelian interaction. As a consequence, the Schwinger confinement
mechanism can be applied to quarks interacting in both the QED
interaction and the QCD interaction, leading to confined QED and QCD
open-string states in (1+1)D, with the composite boson masses
depending on the magnitudes of the QCD and QED coupling constants.
Such a viewpoint is consistent with the QCD string description of
hadrons in the Nambu~\cite{Nam70} and Goto~\cite{Nam70} string model,
the string fragmentation models of particle production of Bjorken,
Casher, Kogut, and Susskind~\cite{Cas74}, the classical yo-yo string
model~\cite{Art74}, the Lund model~\cite{And83}, the Abelian
projection model~\cite{tHo80,Bel79}, and the Abelian dominance model
\cite{Sei07,Suz08,Sug21}.

In a phenomenological analysis, we inquire whether the
phenomenological open-string model of QCD and QED mesons in (1+1)D can
be the idealization of a flux tube showing up as a bound and confined
physical meson in (3+1)D.  In such a phenomenological open-string
model, we need an important relationship to ensure that the boson mass
calculated in the lower (1+1)D can properly represent the mass of a
physical boson in (3+1)D.  The open string in (1+1)D can describe a
physical meson in (3+1)D if the structure of the flux tube is properly
taken into account.  This can be achieved by relating the coupling
constant in (1+1)D with the coupling constant in (3+1)D and the flux
tube radius $R_T$~\cite{Won09,Won10,Won22a}.  Using such a
relationship, we find that that $\pi^0, \eta$, and $\eta'$ can be
adequately described as open-string $q\bar q$ QCD mesons.  By
extrapolating into the $q\bar q$ QED sector in which a quark and an
antiquark interact with the QED interaction, we find an open-string
isoscalar $I(J^\pi)$=$0(0^-)$ QED meson state at 17.9 MeV and an
isovector $(I(J^\pi)$=$1(0^-), I_3$=0) QED meson state at 36.4~MeV.
The predicted masses of the isoscalar and isovector QED mesons in the
open-string model of the QCD and QED mesons are close to the masses of
the reported X17 and E38 particles observed recently, making them good
candidates for these particles.  Further experimental confirmation of
the X17 and the E38 particles will shed light on the question of quark
confinement for quarks interacting in the Abelian U(1) QED interaction
and will have important implications on the basic properties of quarks
and their interactions.

\vspace*{0.5cm}
{\bf Acknowledgments}
\vspace*{0.3cm}

{{The author wishes} 
to thank A. Koshelkin, I-Yang Lee, Y. Jack  Ng, 
Kh.\ Abraamyan, H. N. da Luz, Che-Ming Ko, Siu Ah Chin, P. Adsley, Gang Wang, Yunshan Cheng, H. Holme,  
\linebreak  S. Sorensen, T. Awes, and N. Novitzky for helpful discussions and
communications.  The research was {supported in part by the Division of
Nuclear Physics, U.S. Department of Energy under Contract
DE-AC05-00OR22725 with UT-Battelle, LLC.} 
}

\end{document}